\documentclass[twocolumn,floatfix,superscriptaddress,aps,longbibliography]{revtex4-1}

\usepackage{graphicx}
\usepackage{color}
\usepackage{amsmath,amssymb,amstext} 
\usepackage{float}  
\usepackage[caption=false]{subfig}

\newcommand{\bb}[1]{\mathbf{{#1}}}

\begin{document}

\title{Transient dynamics of a superconducting nonlinear oscillator}

\author{P. Bhupathi}
\thanks{Present Address: Department of Physics, California Institute of Technology, Pasadena, CA 91125, USA}
\affiliation{Department of Physics, Syracuse University, Syracuse, New York 13244-1130}
\author{Peter Groszkowski}
\thanks{Present Address: Department of Physics and Astronomy, Northwestern University, Evanston, IL 60208, USA}
\affiliation{Institute for Quantum Computing and Department of Physics and Astronomy, University of
Waterloo, 200 University Ave W, Waterloo, ON, N2L 3G1, Canada}
\author{M.P. DeFeo}
\affiliation{Department of Physics, Syracuse University, Syracuse, New York 13244-1130}
\author{Matthew Ware}
\affiliation{Department of Physics, Syracuse University, Syracuse, New York 13244-1130}
\author{Frank K. Wilhelm}
\affiliation{Institute for Quantum Computing and Department of Physics and Astronomy, University of
Waterloo, 200 University Ave W, Waterloo, ON, N2L 3G1, Canada}
\affiliation{Theoretical Physics, Saarland University, 66123 Saarbr\"ucken, Germany}
\author{B.L.T. Plourde}
\affiliation{Department of Physics, Syracuse University, Syracuse, New York 13244-1130}

\date{\today}


\begin{abstract}
We investigate the transient dynamics of a lumped-element oscillator based on a dc superconducting quantum interference device (SQUID). The SQUID is shunted with a capacitor forming a nonlinear oscillator with resonance frequency in the range of several GHz. The resonance frequency is varied by tuning the Josephson inductance of the SQUID with on-chip flux lines. We report  measurements of decaying oscillations in the time domain following a brief excitation with a microwave pulse. The nonlinearity of the SQUID oscillator is probed by observing the ringdown response for different excitation amplitudes while the SQUID potential is varied by adjusting the flux bias. Simulations are performed on a model circuit by numerically solving the corresponding Langevin equations incorporating the SQUID potential at the experimental temperature and using parameters obtained from separate measurements characterizing the SQUID oscillator.  Simulations are in good agreement with the experimental observations of the ringdowns as a function of applied magnetic flux and pulse amplitude. We observe a crossover between the occurrence of ringdowns close to resonance and adiabatic following at larger detuning from the resonance. We also discuss the occurrence of phase jumps at large amplitude drive. Finally, we briefly outline prospects for a readout scheme for superconducting flux qubits based on the discrimination between ringdown signals for different levels of magnetic flux coupled to the SQUID.
\end{abstract}

\maketitle

\section{Introduction}
\label{sec:Introduction}
Superconducting circuits composed of Josephson junctions have been the subject of intense research for the past few decades for their importance in understanding the fundamental aspects of quantum mechanics as well as for their potential application towards quantum information processing and computing \cite{leggett2002,clarke2008}. These $\mu$m-sized devices have been shown to exhibit macroscopic quantum tunneling \cite{devoret1985,balestro2003}, quantized energy levels \cite{martinis1985} and superposition of states in a quantum bit (qubit) \cite{nakamura1997,vanderwal2000,Martinis02}. A central application of Josephson devices, classical or quantum, \cite{SQUID-Handbook} is measurement. There have been many advances in utilizing Josephson devices such as a dc superconducting quantum interference device (SQUID) for qubit readout \cite{chiorescu2004, lupascu2007,vion2002, claudon2007}. In these experiments, the SQUID forms part of a resonant oscillator circuit that is coupled to a qubit. Some of these Josephson devices configured as amplifiers have approached the quantum limit in noise performance \cite{siddiqi2005,vijay2009,mueck2010,Hover12,Yurke89,Castellanos08,Hatridge11,Bergeal2010,Chen11,Govia12}. All of these applications of Josephson junctions depend at some level on the nonlinearity of the junction response.

In this paper we investigate the temporal dynamics of a nonlinear SQUID resonant circuit.  Studies of nonlinear oscillator dynamics under continuous excitation have been done previously \cite{vijay2009,marchese2006classical,claudon2004coherent,lisenfeld2007temperature,gronbech2005rabi}, and, in fact, many types of superconducting qubits, such as the transmon \cite{Koch07} or the phase qubit \cite{Martinis02}, are nonlinear oscillators that are typically driven with resonant pulses. However, we are not aware of any experimental or theoretical work to date on the transient dynamics of nonlinear oscillators under pulsed excitation in the time domain. Here we present time domain measurements of the decaying voltage oscillations from the SQUID oscillator after a brief excitation. The SQUID potential, and hence the resonance frequency, can be tuned by changing either the bias flux or bias current; here we focus on the variation with respect to flux, while no dc bias current is applied.

The paper is organized as follows. We start with a brief theoretical background of SQUID oscillators in Sec.~\ref{sec:TheoreticalIntro}. In Sec.~\ref{sec:ExperimentalSetup} we describe the fabrication of a lumped element SQUID oscillator and experimental measurement scheme. Measurements in the frequency and time domains are presented in Sec. \ref{sec:MeasOfSquidOscillator}. A model of the electrical circuit is presented in Sec.~\ref{sec:ModelAndSimulations}, which is then used to derive the equations of motion of the full system. These equations are then reduced, accounting for the physical parameters used, and after incorporating thermal effects, are solved numerically to obtain the free evolution of the system for different conditions corresponding to the experiment. The simulated ringdowns in the time domain as a function of flux bias and pulse amplitude are in good agreement with the observations, as is shown in Sec.~\ref{sec:theory_ringdowns}.  Finally, in Sec.~\ref{sec:Applications}, a scheme for using ringdown oscillations of a SQUID oscillator to read out a flux qubit is briefly discussed, followed by our conclusions in Sec. \ref{sec:Conclusions}.

\section{Theoretical background}
\label{sec:TheoreticalIntro}

At the heart of all these investigations is the Josephson junction, which behaves as a nonlinear LC-oscillator characterized by the plasma resonance \cite{Tinkham96,SQUID-Handbook}, $\omega_{p} = \sqrt{2\pi I_{0}/\Phi_{0} C_{J}}$ , where, $I_{0}$ is the critical current of the Josephson junction in parallel with its self-capacitance $C_{J}$ and $\Phi_{0} \equiv h/2e$ is the magnetic flux quantum. For typical parameters of the fabricated junctions, $\omega_{p}/2\pi$ is of the order of $100\,{\rm GHz}$. When shunted by a large external capacitance however, the resonance frequency can be lowered to a few GHz for ease of performing experiments with an oscillator for coupling to a qubit or to fabricate a qubit itself. 


In this paper we consider a dc SQUID, which has two identical junctions in parallel, symmetrically placed on a superconducting loop. The dynamics of such a SQUID, as shown in a circuit schematic of Fig.~\ref{fig:schematic-fab}, can be described by a two-dimensional anharmonic potential $U(\varphi_{+},\varphi_{-})$ given by \cite{SQUID-Handbook,lefevre-sequin1992}
\begin{equation}
    \frac {U} {2 E_{J}} = -i_{b} \varphi_{+} - \cos \varphi_{+} \cos \varphi_{-}   + \frac {1} {\beta} \left( \varphi_{-} - \pi f_{s} \right)^{2} 
    \label{eq:squid-pot}
\end{equation}
where $E_{J}=I_{0}\Phi_{0}/2\pi$ is the Josephson coupling energy normalizing $U$, $L_{J0} = \Phi_{0}/2\pi I_{0}$ is the Josephson inductance of each junction, $\beta=2 \pi I_{0} L_{g}/\Phi_{0} = L_{g}/L_{J0}$ is the screening parameter of the SQUID with geometric loop inductance $L_{g}$, $i_{b}=I_{b}/2I_{0}$ is the bias current normalized by the critical current, and $f_{s}=\Phi_{s}/\Phi_{0}$ is the normalized applied flux. $\varphi_{+} = (\varphi_{A}+\varphi_{B})/2$ and $\varphi_{-} = (\varphi_{A}-\varphi_{B})/2$ are the two independent degrees of freedom with $\varphi_{A}$ and $\varphi_{B}$ corresponding to the phase differences across each of the two Josephson junctions (see Fig.~\ref{fig:circuit-model}(a) and Appendix~\ref{sec:EquationsOfMotion} for a more complete description of the SQUID potential energy and the rest of the circuit). 
The sum of the phases across the junctions (external mode), $\varphi_{+}$, couples to the current through the SQUID, while the difference of the phases (internal mode), $\varphi_{-}$, couples to the magnetic flux applied to the SQUID. The oscillator can be resonantly excited by applying a short alternating current pulse to the SQUID. The pulse perturbs the potential minimum of the SQUID, giving rise to oscillations of the phase particle about the minimum that decay at the characteristic frequency of the oscillator. The ringdown motion is mediated via the external mode of the oscillator and can be detected as a voltage oscillation across the SQUID. 

\section{Experimental setup}
\label{sec:ExperimentalSetup}
\begin{figure}[h]
\centering
  \includegraphics[width=3.35in]{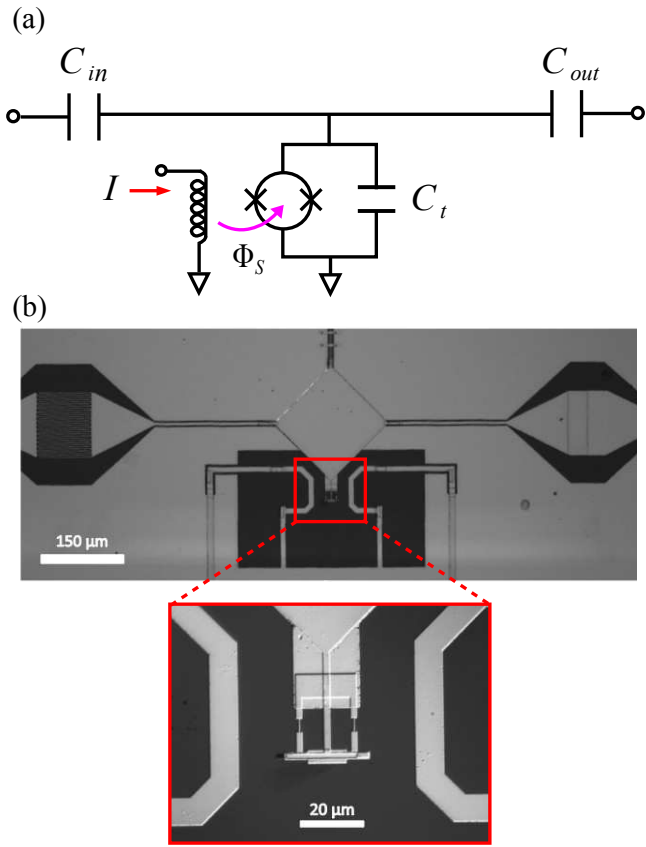}
  \caption{(Color online) (a) Circuit schematic showing the input/output coupling capacitors and the 
      SQUID oscillator. (b) Optical image of the fabricated circuit with a zoomed-in view of the SQUID with on-chip flux lines.
\label{fig:schematic-fab}}
\end{figure}

We investigate a lumped-element microwave oscillator circuit consisting of a dc SQUID shunted by a capacitor formed from superconducting layers. A circuit schematic and an optical micrograph of the device are shown in Fig.~\ref{fig:schematic-fab}. A microwave feedline with on-chip capacitors couples signals into and out of the oscillator. Adjusting the bias flux $\Phi_{s}$ as shown in the figure modulates the Josephson inductance of the SQUID, thus varying the resonance frequency of the SQUID oscillator.

Our devices are fabricated in a 5-layer process on an oxidized Si wafer. The initial four layers are patterned using photolithography, while the final layer, consisting of the SQUID junctions, is patterned by electron-beam lithography. The ground plane is formed from a $120\,{\rm nm}$-thick Al layer. The dielectric layer on top of the ground plane is a $150\,{\rm nm}$-thick SiO$_2$ film deposited by plasma enhanced chemical vapor deposition (PECVD). The SiO$_2$ forms the dielectric for the parallel plate shunting capacitor $C_{t}$ and the output coupling capacitor $C_{\rm out}$. The input coupling capacitor $C_{\rm in}$ (Fig.~\ref{fig:schematic-fab}) is interdigitated and is formed along with the microwave feedline and the top layer of the parallel plate capacitors in a $200\,{\rm nm}$-thick Al film. The estimated value of $C_{\rm in}$, based on the fabricated finger dimensions and using the effective dielectric constant of a microstrip line in the standard expression for the interdigitated capacitor \cite{yoon2000,gupta96}, is $0.15\,{\rm pF}$. The parallel plate capacitors $C_{\rm out}$ and $C_{t}$ are designed to be $1.3\,{\rm pF}$ and $6.8\,{\rm pF}$, respectively, based on geometry and film parameters, although there could be significant variation because of our uncertainty in the dielectric properties of the SiO$_2$ film. Vias are etched through the SiO$_2$ layer so that the bias lines and one end of the SQUID contact the ground plane. The SQUID loop is formed so that it is coupled symmetrically between the top and bottom plates of the shunt capacitor [Fig.~\ref{fig:schematic-fab}(b)]. The geometric inductanace $L_{g}$ of the SQUID is calculated to be $43\,{\rm pH}$, from FastHenry simulations of the loop dimension $(18\times18\,{\rm \mu m^2})$.

The SQUID junctions are Al-AlO$_x$-Al, formed by a standard double-angle shadow-evaporation method  \cite{dolan1977} in a dedicated electron-beam evaporation chamber equipped with {\it in situ} Ar ion milling to ensure superconducting contacts between the SQUID layer and the junctions. The junctions are sub-micron in size, $530\times160\,{\rm nm^2}$, with a junction capacitance estimated to be $10\,{\rm fF}$ \cite{deppe2004}. The critical current of each  junction was estimated to be $0.4\,{\rm \mu}$A by measuring the normal state resistance of a nominally identical junction $(684\,{\rm \Omega})$ and based on our previous characterizations of similar sized junctions \cite{DeFeo2010}. The resonance frequency of the oscillator was designed to be at $3\,{\rm GHz}$.

\begin{figure}[h]
\centering
  \includegraphics[width=0.36\textwidth]{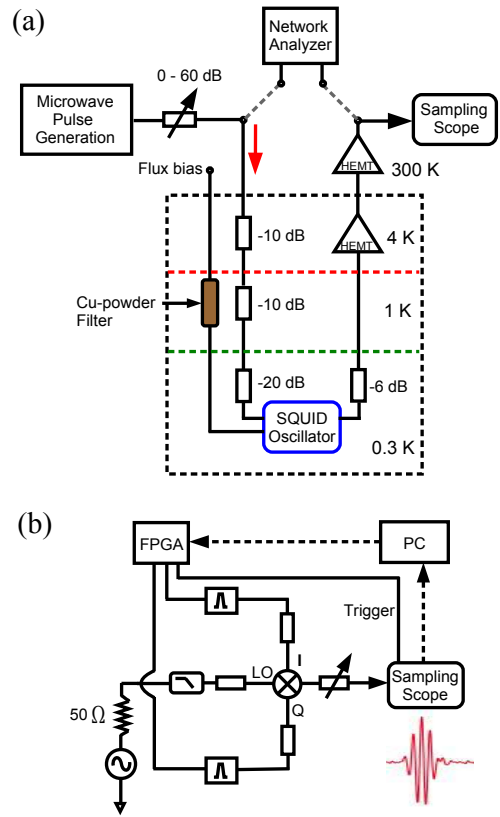}
  \caption{(Color online) (a) Schematic of general measurement setup. The network analyzer is used to measure the frequency response  while the pulse generation setup (expanded in (b)) is used to measure the ringdowns in the time domain. The two types of measurements are separate and, as indicated by dashed lines in (a), the network analyzer is never connected simultaneously with the pulse generation setup. The SQUID oscillator device shown as a blue box is displayed in Fig. 1.}
  \label{fig:measurement}
\end{figure}

Measurements are performed in a $^3$He cryostat with a base temperature of $300\,{\rm mK}$. A schematic of the measurement setup is shown in Fig.~\ref{fig:measurement}(a). A vector network analyzer is used to characterize the resonance frequency, while the transient dynamics are studied using a custom-built GHz DAC (digital to analog converter) and a $20\,{\rm GHz}$ sampling oscilloscope. The sample chip is wire-bonded to a Cu stripline microwave board and enclosed in an aluminum box for magnetic shielding, which is anchored to the cold plate of the $^3$He cryostat. The drive-line to the SQUID oscillator is a lossy stainless steel semirigid coaxial cable with attenuators heat sunk at various stages of the cryostat to minimize noise from  room temperature. The transmitted signals at the output are amplified by two High Electron Mobility Transistor (HEMT) amplifiers: one at the $4\,{\rm K}$ stage of the cryostat and another at room temperature with a combined gain of $70\,{\rm dB}$. A $6\,{\rm dB}$ attenuator is used at the output of the oscillator for $50\,{\rm \Omega}$ impedance matching to the input of the cryogenic HEMT amplifier. The dc biasing lines have copper powder filters anchored at the $1\,{\rm K}$ stage of the cryostat. A cryogenic $\mu$-metal can surrounds the vacuum can of the cryostat to shield the SQUID from external magnetic fields. 

Generation of the microwave pulses is achieved by employing a FPGA based DAC board, which has been built based on the designs from UCSB \cite{martinisDAC}. A schematic of the pulse generation setup is shown in Fig.~\ref{fig:measurement}(b). A nanosecond waveform is generated digitally from the FPGA controlled from a computer. The waveform is passed through Gaussian filters and attenuators before being mixed with a resonant carrier tone at the IQ-mixer, producing a short microwave burst at the output port of the mixer. Different amplitudes of the microwave pulse are achieved by varying the attenuation after the output stage of the mixer with a step attenuator. A typical room temperature trace of the microwave bursts generated by this setup and corresponding time domain measurement is discussed in the next section.

\section{Measurements of the SQUID Oscillator}
\label{sec:MeasOfSquidOscillator}
The resonance frequency of the SQUID oscillator was characterized as a function of the applied flux at zero bias current ($I_{b}=0$).  A network analyzer supplied a weak signal with a typical power of $-125\,{\rm dBm}$ to the input of the SQUID oscillator and the 2-port complex transmission parameter, $S_{21}(f)$, was measured. Although, we measured the complex quantity, $S_{21}$, the plots that we present only show the magnitude, $|S_{21}|$.
Multiple $|S_{21}|$ traces were recorded while stepping through the flux applied to the SQUID. Figure~\ref{fig:SQUID-mod} shows a density plot of the flux-modulated resonance frequency, periodic in $\Phi_{0}$, with the color scaling (darker blue) region indicating the highest magnitude of $|S_{21}|$ and the flux axis scaled in units of $\Phi_{0}$. 
A phenomenological function of the form $f(V_{\Phi s})=\sqrt{a \mid \cos\left( b V_{\Phi s}+c\right) \mid}$, where $V_{\Phi s}$ is the flux bias voltage, is used to fit the flux modulation of the resonance peaks over two periods. The fit parameters $a$, $b$, $c$ are used to scale the flux axis and fit the frequency values, so that the fit curve (dashed gray line) can be plotted on top of the scaled resonance modulation data, as shown in Fig.~\ref{fig:SQUID-mod}. The highest frequency at integer $\Phi_{0}$ occurs at  $3.2\,{\rm GHz}$. As expected, for $\beta \ll 1$, the critical current, and hence the resonance frequency modulate to near zero. The observed quality factor of the resonance is $\sim7$, consistent with the estimated value for the output coupling capacitor and the chip parameters as discussed in the previous Section~\ref{sec:ExperimentalSetup}. In the following section, we describe a circuit model used to fit the measured $|S_{21}|$ traces as a function of applied flux by calculating the flux-dependent Josephson inductance based on the SQUID potential.

\subsection{Transmission through a SQUID oscillator with asymmetric coupling}
\begin{figure}[h]
\centering
  \includegraphics[width=0.36\textwidth]{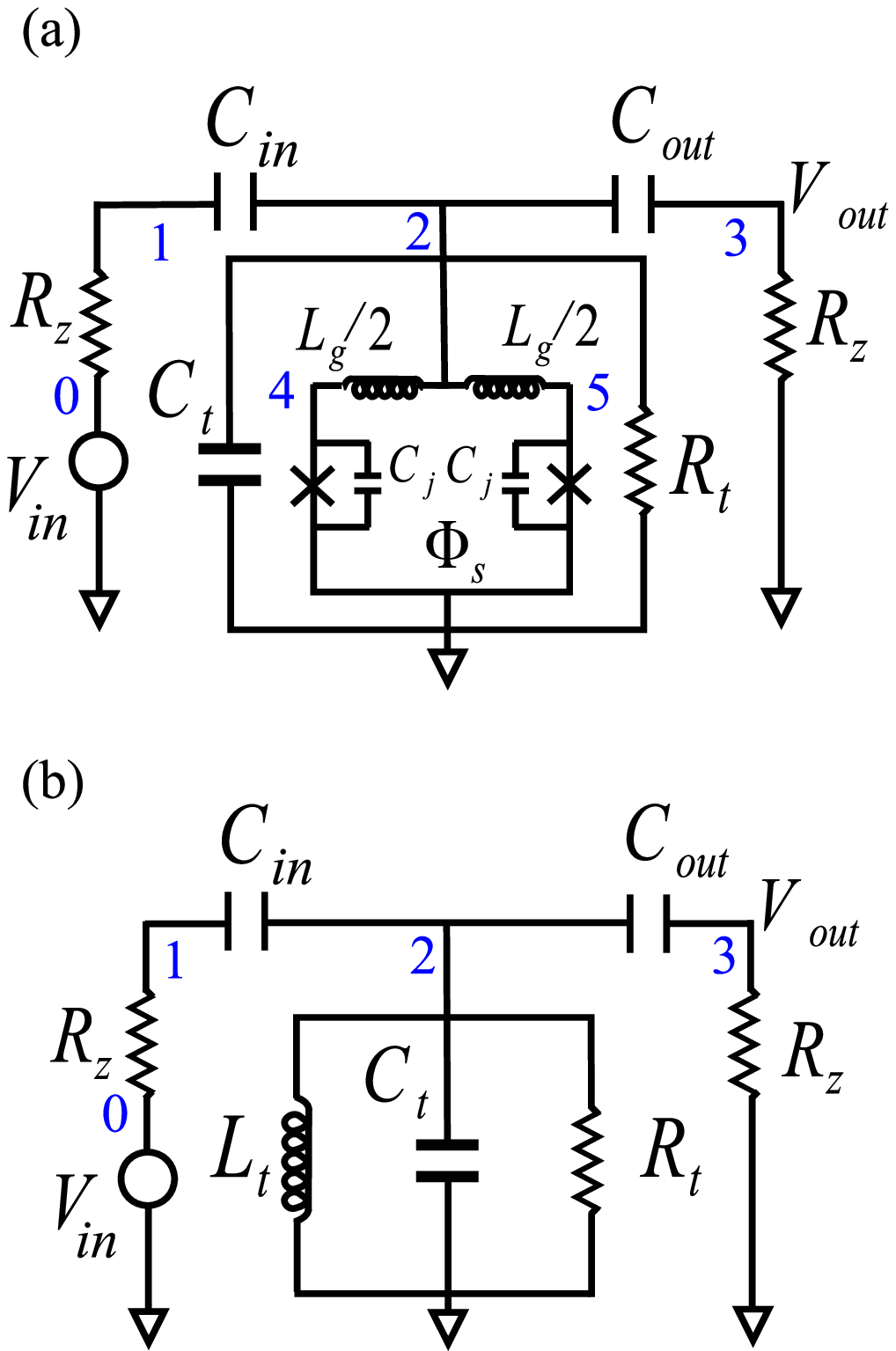}
  \caption{(Color online) (a) Full circuit model of the SQUID oscillator of Fig.~\ref{fig:schematic-fab} and (b) reduced circuit, valid in the limit of small amplitudes when the Josephson junction responds linearly. In this regime the SQUID is treated as an effective, flux-dependent inductance $L_{t}$. In both (a) and (b), the blue numbers represent node labels.
\label{fig:circuit-model}}
\end{figure}

A full circuit model is shown in Fig.~\ref{fig:circuit-model}(a), where a Josephson junction, with parallel capacitance $C_{J}$, is in series with a geometric inductance $L_{g}/2$, symmetrically positioned on each side of the SQUID.  The external shunt capacitance is labeled as $C_{t}$ and the internal dissipation in the SQUID oscillator due, for example, to losses in the shunt capacitor, is depicted as $R_{t}$. $C_{\rm in} (C_{\rm out})$ is the input (output) coupling capacitor and $R_{z}$ is the characteristic impedance at the input and output. Neglecting the effects of the small junction capacitances $C_{J}$ and noting that the dc bias current $I_{b}=0$, at low-amplitude drive, such that the Josephson junction is nearly linear, we can describe the SQUID as an effective inductance \footnote{In the linear regime of the Josephson junctions, this expression can be derived by standard circuit analysis treating individual junction inductances as $L_{J0}/\cos (\varphi)$, with $\varphi$ representing the phase across each junction, or alternatively by expanding the potential energy in Eq.~(\ref{eq:squid-pot}) around the minima of $\varphi_{\pm}$. The term $L_g/4$ comes from the fact that in the case of non-zero geometric inductance $L_{g}$, the voltage across the SQUID is not $\phi_{0} \dot  \varphi_{+}$, but instead $ \phi_{0} \dot \varphi_{2}$ (as can be seen from Fig.~\ref{fig:circuit-model}). Yet another, equivalent way of calculating $L_{t}$, would be to expand the effective potential energy discussed in the Appendix~\ref{sec:EquationsOfMotion} around the minimum of $\varphi_{2}$.}


\begin{equation}
    L_{t} = \frac {L_{J0}} { 2 \cos \varphi_{-}^{\rm min} } + \frac{L_{g}}{4}.
\label{eq:squid-Lmod}
\end{equation}
Here $\varphi_{-}^{\rm min}$ represents the steady state value of $\varphi_{-}$, which for a given value of the applied flux $f_s$, we can calculate numerically by minimizing the potential energy described by Eq.~(\ref{eq:squid-pot}).


Using Eq.~(\ref{eq:squid-Lmod}) we can derive a general expression for $S_{21}$ for the effective parallel LCR tank circuit with asymmetric input and output coupling capacitors. A similar analysis with symmetric coupling capacitors was performed in Ref.~\cite{oconnell2008}. We define $S_{21}=2V_{\rm out}/V_{\rm in}$ such that a matched load of $R_{z}=50\,\Omega$ corresponds to full transmission, $S_{21}=1$ if the input were connected directly to the output. Then, taking the notation $m\parallel n$ to represent the parallel impedance between impedance elements $m$ and $n$, we arrive at
\begin{equation}
    S_{21}(\omega, L_{t}) = \frac{2V_{\rm out}}{V_{\rm in}} = 2 \frac{R_{z}} {Z_{\rm out}} \frac {(Z_{t}\parallel Z_{\rm out})} {(Z_{t}\parallel Z_{\rm out}) + Z_{\rm in}},
\label{eq:squid-s21}
\end{equation}
 where $Z_{t}$ is the impedance of the parallel LCR tank circuit,
\begin{equation}
    Z_{t} = \left(\frac{1}{R_{t}} + \frac {1}{i \omega L_{t}} + i \omega C_{t} \right)^{-1}
\label{eq:z-lcr}
\end{equation}
and $ Z_{\rm in} = \left( R_{z} + \frac {1}{i \omega C_{\rm in}} \right)$, $ Z_{\rm out} = \left( R_{z} + \frac {1}{i \omega C_{\rm out}} \right)$ are the input and output impedances respectively.

Equation (\ref{eq:squid-s21}) was used to fit the measured $S_{21}$ traces at each flux bias value. First, the measured $S_{21}$ traces were scaled by the $S_{21}$ transmission at low temperature that was measured on a separate cooldown using a direct coaxial connection in place of the oscillator chip. This scaling of measured $S_{21}$ of the SQUID oscillator resonance by the effective low-temperature baseline takes into account the temperature dependence in the system transmission.

\begin{figure}[h]
\centering
 \includegraphics[width=3.35in]{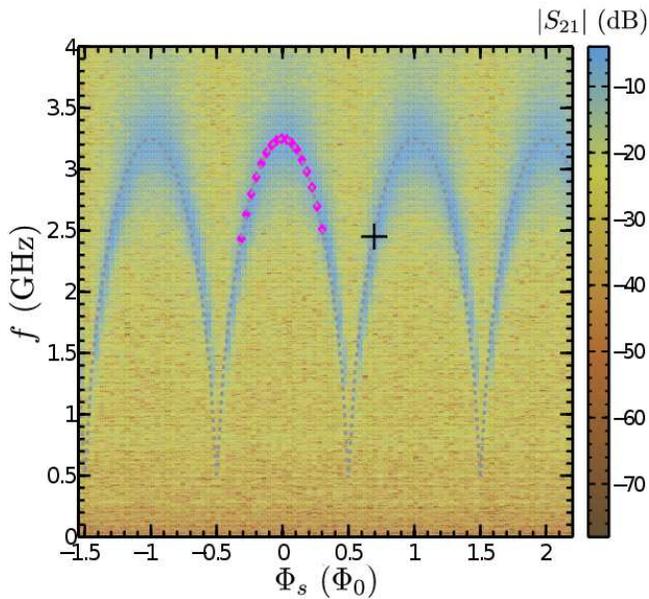}
  \caption{(Color online) Density plot of $|S_{21}|$ vs. flux and frequency of the SQUID oscillator as measured from a network analyzer at $-125\,{\rm dBm}$ power at the input of the SQUID oscillator chip at $300\,{\rm mK}$. The dashed line and magenta symbols are from fits to the SQUID modulation as described in the text. The marker in black indicates the bias point where the pulsed measurements were taken.
\label{fig:SQUID-mod}}
\end{figure}

We fit the $S_{21}(f)$ data by fixing $I_{0}=0.4\,{\rm \mu}$A, $C_{\rm in}=0.15\,{\rm pF}$ and $\beta=0.05$ and varying $C_{t}$ and $R_{t}$. The estimates for the fixed parameters were explained earlier in Section~\ref{sec:ExperimentalSetup}. Measured $S_{21}$ curves between $\pm0.3\,\Phi_{0}$ were fit simultaneously with the same fixed parameters and the best fit parameters extracted for $C_{t}$ and $R_{t}$ were $5.1\,{\rm pF}$ and $265\,{\rm \Omega}$ respectively. $C_{t}$ is in reasonable agreement with our estimate based on fabrication parameters described in Section~\ref{sec:ExperimentalSetup}. We can extract a loss tangent $\tan \delta$ for SiO$_2$ from the equivalent resistance  $R_{t}$ at a frequency of $2.5\,{\rm GHz}$ as $1/ \omega C_{t} R_{t} \sim 4 \times 10^{-2}$. We note that this is roughly an order of magnitude larger than what was measured in Ref.~\cite{oconnell2008} for PECVD-deposited SiO$_2$ and this difference could be due to variations in deposition conditions and film quality. Also, a precise determination of the internal loss in our capacitor is difficult with these measurements because the large output coupling strength limits the total oscillator quality factor in our circuit.  
The resonance peaks from the circuit model fits are plotted in Fig.~\ref{fig:SQUID-mod} as solid magenta symbols.
The extracted fit parameters $C_{t}$ and $R_{t}$ are further used in the time-domain simulations described in Section~\ref{sec:ModelAndSimulations}.

\begin{figure}[tp]
    \begin{center}
        \includegraphics[width=0.36\textwidth]{{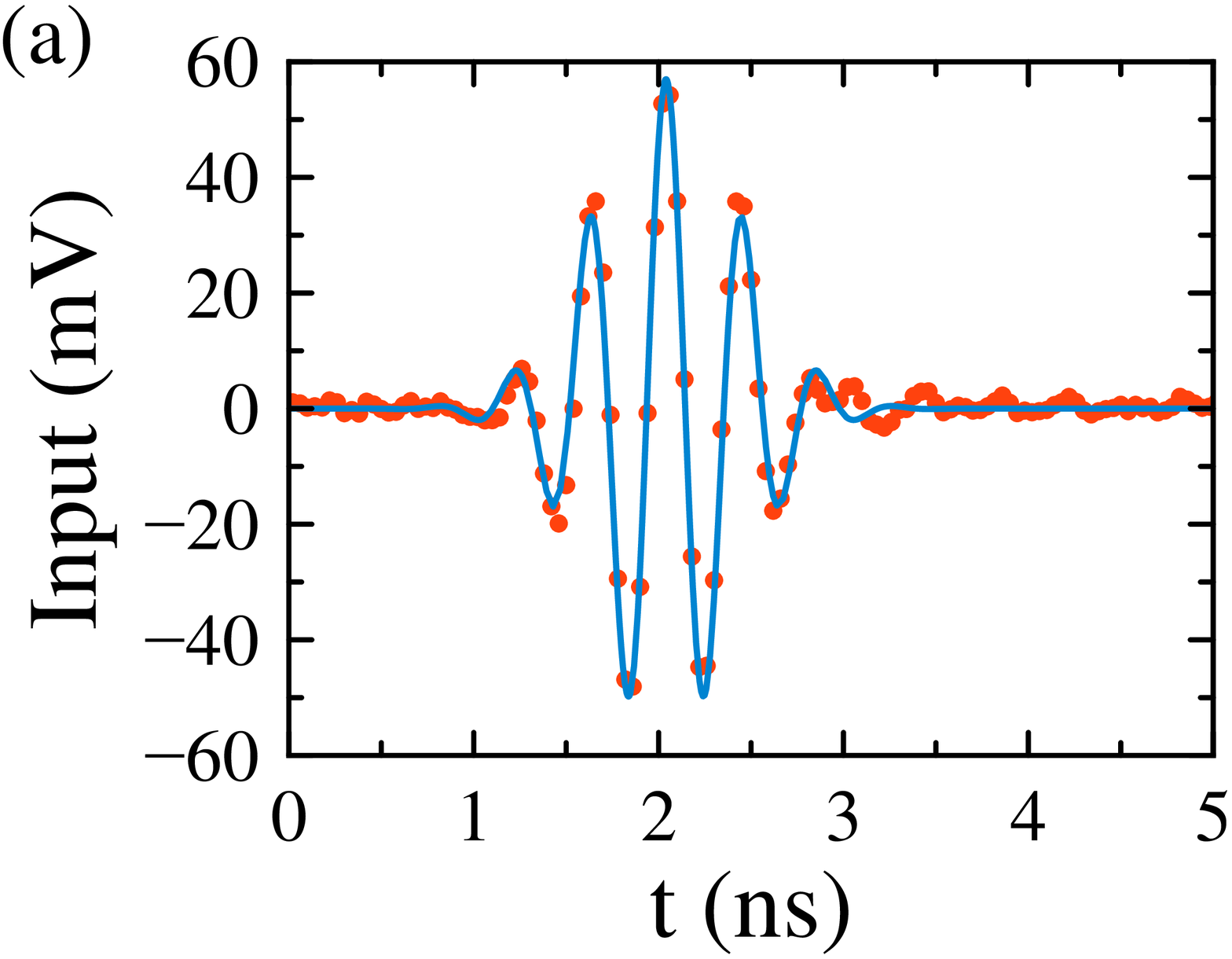}} 
        \includegraphics[width=0.36\textwidth]{{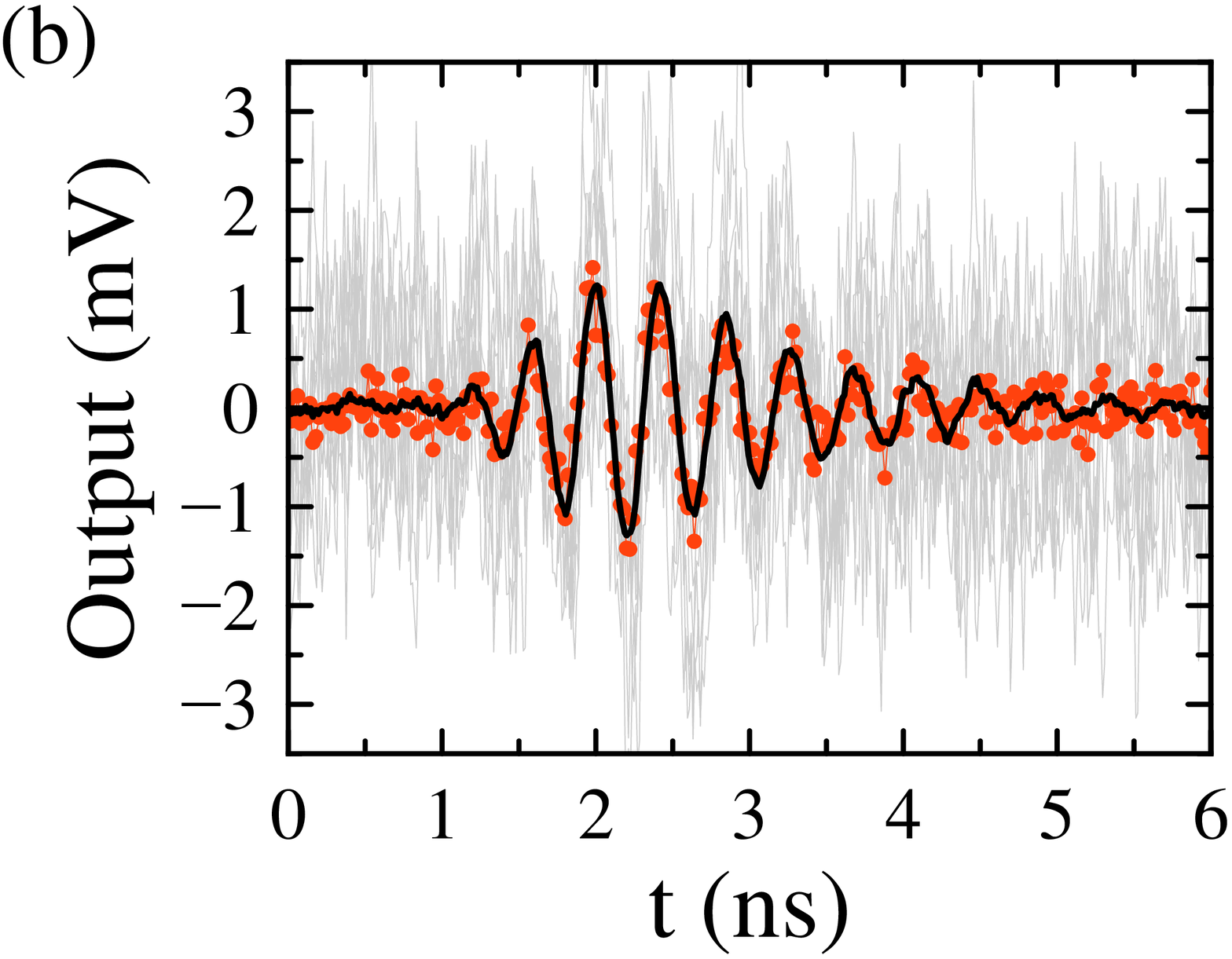}} 
    \end{center}
    \caption{(Color online) (a) Input pulse as measured at the top of the cryostat. The red dots represent the measured data, while the blue curve is the fit used as the input voltage in the simulations. (b) Example measured output voltage ringdown from the SQUID oscillator. Once again, the red dots were obtained directly from experimental data, while the gray curves in the background are a subset of the realizations obtained from the simulations of the circuit. The average of these realizations is shown in black. }
    \label{fig:single-ringdown}
\end{figure}

Time-domain measurements were performed by applying a short microwave burst to the SQUID oscillator that was flux biased at $\Phi_{s} = 0.3\,\Phi_{0}$, where the resonance frequency, and hence the ringdown waveforms, have high flux sensitivity. An example microwave pulse generated by the GHz DAC is shown in Fig.~\ref{fig:single-ringdown}(a) along with the corresponding theoretical fit used for simulations of the ringdowns described in Section~\ref{sec:ModelAndSimulations}. The voltage signal at the output of the SQUID oscillator is shown in Fig.~\ref{fig:single-ringdown}(b) for the case when the SQUID is flux biased at $0.3\,\Phi_{0}$, corresponding to a resonance frequency of 2.4 GHz. A single experimental ringdown trace shown in the figure is an average of 1000 traces on the sampling oscilloscope. The carrier frequency of the microwave burst is $2.4\,{\rm GHz}$, to be on resonance when the SQUID oscillator is biased at $0.3\,\Phi_{0}$

The SQUID ringdown response to various amplitudes of the drive pulse was mapped out by varying the attenuation on the drive line of the SQUID oscillator. The attenuation was varied in steps of 1 dB from 0 to 40 dB at the top of the cryostat. At a setting of 0 dB, that is, with no extra attenuation other than the fixed attenuation inside the cryostat, the measured peak amplitude of the burst at the top of the cryostat was 50 mV. 
Figure~\ref{fig:single-ringdown}(b) displays experimental and theoretical ringdown traces, to be described in more detail in the next section, from the SQUID oscillator when the pulse amplitude driving the SQUID is much smaller than its critical current. The simulated ringdown is shown as a black line while the experimental data is in solid circles.

\section{Model and Simulations}
\label{sec:ModelAndSimulations}

In Sec.~\ref{sec:MeasOfSquidOscillator}, where the calibration procedure was described, and where only small drive amplitudes were considered, we treated the Josephson junctions as linear elements, and hence the SQUID as a simple effective inductance. In this section we consider a more complete picture of the full circuit, shown in Fig.~\ref{fig:circuit-model}(a), which accounts for the nonlinearity that is relevant when the applied pulse amplitude is high. 
In Appendix~\ref{sec:EquationsOfMotion} we first write down the equations of motion of the full system in terms of five degrees of freedom. Since in our case, the shunt capacitance $C_{t}$ is much larger than the junction capacitances $C_{J}$, and the Josephson inductance of the junctions $L_{J0}$ is in turn much larger than the geometric inductance of the SQUID, $L_{g}$ (a condition we can write as  $\beta = L_{g} / L_{J0} \ll 1$), we can eliminate the fastest degrees of freedom corresponding to nodes $4$ and $5$ in Fig.~\ref{fig:circuit-model}(a). Doing so, leaves us with a set of differential equations that govern the behavior of our circuit, of only three degrees of freedom that treat the SQUID potential energy as one-dimensional.
Furthermore, in Appendix~\ref{sec:EquationsOfMotion}, we also discuss how, using the thermodynamic dissipation-fluctuation theorem \cite{nyquist1928thermal}, we can include the effects of noise due to the non-zero temperature of the system. This stochastic noise turns out to play a crucial role in reproducing the experimental behavior when high amplitude input pulses are considered. Combining all of these factors results in the effective equations of motion (in units of current), which can be written using vector notation as
\begin{align}
    \phi_{0} \bb{C} \ddot{\vec{\varphi}} +  \phi_{0} \bb{R^{-1}} \dot{ \vec{\varphi }} + \frac{1}{\phi_{0}} \vec{ \nabla}_{\varphi} U_{\rm eff} + \bb{N} \vec{n} + \vec{I}_{\rm{dr}} = 0.
    \label{eq:reducedSystem2Vec1}
\end{align} 
Here $\vec{\varphi}=(\varphi_{1}, \varphi_{2}, \varphi_{3})^{T}$, $ \vec{ \nabla}_{\varphi}  = (\partial / \partial \varphi_{1}, \partial / \partial \varphi_{2}, \partial / \partial \varphi_{3})^{T} $,  $\vec{I}_{\rm{dr}}=(- V_{\rm{in}}/R_{z}, 0, 0)^{T}$, $\vec{n}=(n_{1}, n_{2}, n_{3})^{T}$ and $\phi_{0}=\Phi_{0}/2\pi$. Each variable $\varphi_{i}$, represents the superconducting phase at node $i$, with the corresponding voltage defined as $\phi_{0} \dot \varphi_{i}$. The thermal noise in the circuit is modeled by including a current noise source of strength $\sqrt{ \frac{2 k_{B} T }{ R_i}} n_{i}$ in parallel with each resistor $R_{i}$. We take $k_{B}$ to represent the Boltzmann constant, T the temperature and each $n_{i}$, a normally distributed random variable,
which satisfies
\begin{align}
    \langle n_{i}(t) \rangle =&  0 \label{eq:noise2aave} \\
    \langle n_{i}(t) n_{j}(t') \rangle =&  \delta(t-t') \delta_{i,j}.  \label{eq:noise2a}
\end{align}
In our case, taking $C_{\Sigma}=C_{t} + C_{\rm in} + C_{\rm out}$, the matrices corresponding to $\bb{C}$ and $\bb{R^{-1}}$ can be written as
\begin{align}
    \bb{C} =
\begin{pmatrix} 
    C_{\rm{in}}     & -C_{\rm{in}} & 0 \\
    -C_{\rm{in}} & C_{\Sigma} & -C_{\rm{out}} \\
    0 & -C_{\rm{out}} & C_{\rm{out}} 
\end{pmatrix}, 
\bb{R^{-1}}=
\begin{pmatrix} 
    \frac{1}{R_{z}}    & 0 & 0 \\
  0 & \frac{1}{R_{t}}  & 0 \\
   0 & 0 & \frac{1}{R_{z}}  
\end{pmatrix},
\end{align}
and $\bb{N}$ as simply
\begin{align}
    \bb{N} = \sqrt{2 k_{B}T \bb{R^{-1}}}.
\end{align}
Finally, $U_{\rm eff}$ represents the effective (drive free) potential energy of our system, which can be decomposed as $U_{\rm eff}= U_{0} + U_{1}$. The terms $U_{0}$ and $U_{1}$ represent the contributions, up to zeroth and first order in $\beta$ respectively (as explained in Appendix \ref{sec:EquationsOfMotion}), which result in
\begin{align}
    \vec{ \nabla}_{\varphi} U_{0} = 
\begin{pmatrix} 
     0 \\
     2 E_{J} \cos \left( \pi f_{s} \right) \sin \left( \varphi_{2} + \pi f_{s} \right) \\
    0
\end{pmatrix}, 
    \label{eq:dirU0}
\end{align}
\begin{align}
     \vec{ \nabla}_{\varphi} U_{1} =  \beta
\begin{pmatrix} 
     0 \\
     - \frac{E_{J}}{2} \left(\sin(4 \pi f_{s}+2 \varphi_{2})+\sin(2 \varphi_{2}) \right) \\
    0
\end{pmatrix}.
    \label{eq:dirU1}
\end{align}
Eq.~(\ref{eq:reducedSystem2Vec1}) forms a set of stochastic differential equations that we can numerically solve for any $\varphi_{i}$, although each solution only gives us a single realization. Averaging over many such realizations ($500$ in our case) produces a curve that can be directly compared to the experimental data, which we do in Sec.~\ref{sec:theory_ringdowns}. In our simulations we chose the initial conditions that correspond to the system being at rest, near the potential energy minimum, which we can express as $\dot{\vec{\varphi}}(0)= \vec{0}$ and $\vec{\varphi}(0)=(0, -f_s \pi, 0)^{T}$ respectively. We then let the system thermalize by evolving Eq.~\ref{eq:reducedSystem2Vec1} without an external drive present ($\vec{I}_{\rm dr}=\vec{0}$), with only the thermal noise influencing the evolution. In the final step, the input pulse shown in Fig.~\ref{fig:single-ringdown}(a) is applied, which in turn excites the system, leading to ringdown oscillations.


\subsection{Model Limitations}
\label{ssec:ModelLimitations}

In our model, we neglect the internal dissipation associated with each Josephson junction. This is reasonable when the amplitude of the excitation applied to the junctions is smaller than their critical current, as the effective resistance shunting each junction in this case is large enough that its effect on the damping of the junction phase can be neglected. However, in the instances when the driving current exceeds the critical current, each junction experiences a resistance that can be of the order of its normal-state resistance $R_{n}$ \cite{Tinkham96}, which for the SQUID oscillator studied here is $684\,\Omega$. Nevertheless, in our circuit, the dominant source of noise is the $50\,\Omega$ outside load that couples to the SQUID oscillator via $C_{\rm out}$, as can be seen from Fig.~\ref{fig:circuit-model}. When this load is mathematically transformed as impedance parallel with the oscillator over the frequency ranges of the input pulses we apply, its resistive component is never more than $100\,\Omega$, hence a few times smaller than all other sources of noise in the system such that its effect is by far the most dominant.

Furthermore, we neglect any quantum corrections to the noise correlation function and, as is shown in Eq.~(\ref{eq:noise2a}), treat it, just as the rest of the system, fully classically. This is typically a reasonable assumption in the limit of $\hbar \omega \ll 2 k_{B} T$ with $\omega$ being the applied, flux dependent, effective natural frequency of the oscillator circuit. In the case of the experimental parameters used here, this limit is largely satisfied, although in the worst case, when the flux through the SQUID is close to integer multiples of a flux quantum (where the effective natural frequency of the oscillator is largest) we are slowly approaching a case where $\hbar \omega \le 2 k_{B} T$, in particular with $T=0.300\,{\rm K}$ and at $f_{s}=0$, we have $\frac{\hbar \omega}{2 k_{B} T} \sim \frac{1}{4} $. Calculating the leading order correction to the quantum version of the correlation function \cite{callen1951irreversibility,ben1983quantum}, leads to the following adjusted Eq.~(\ref{eq:noise2a})
\begin{align}
    \langle n_{i}(t) n_{j}(t') \rangle =& \left(1 + \frac{1}{48}\right)  \delta(t-t') \delta_{i,j},
    \label{eq:noise2aCorr}
\end{align}
which in turn we can interpret as a factor of $\sim 1/24$ change in the amplitude of the noisy current sources associated with each of the resistors. Given that this is the worst case, and this correction gets smaller as the flux that is threaded through the SQUID shifts away from integer multiple of $\Phi_{0}$, we neglect it in our simulations.


\section{Voltage Ringdowns}
\label{sec:theory_ringdowns}
While we can simulate the evolution of an arbitrary degree of freedom, of particular interest is $ \dot \Phi_{3}=\phi_{0} \dot \varphi_{3}$, as it corresponds to the output voltage, which is precisely what is measured in the experiments. As already discussed, we have two ``knobs'' that can be controlled in a given experimental run; the flux bias $f_{s}$, and the amplitude of the input pulse. The rest of the parameters are fixed at the values described in Sec.~\ref{sec:MeasOfSquidOscillator}. To explore the behavior of our system better, it is therefore instructive to vary one of these control knobs while keeping the other constant. 


\begin{figure}[tp]
  \captionsetup[subfigure]{position=top,singlelinecheck=off,topadjust=-5pt,justification=raggedright}
  \centering
  \subfloat[]{            \includegraphics[width=0.36\textwidth]{{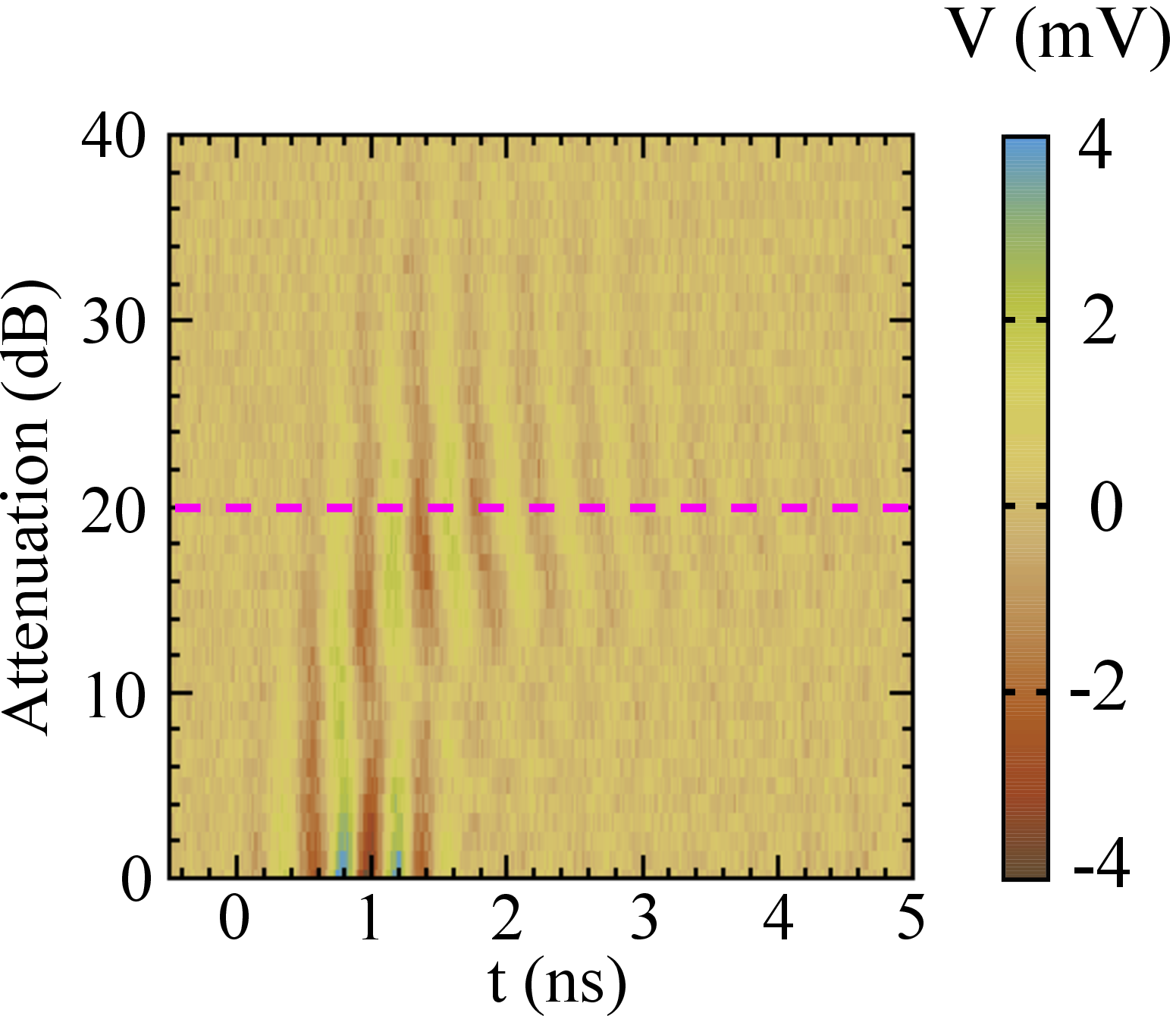}}  } \\
  \vspace{-0.4cm}
  \subfloat[]{            \includegraphics[width=0.36\textwidth]{{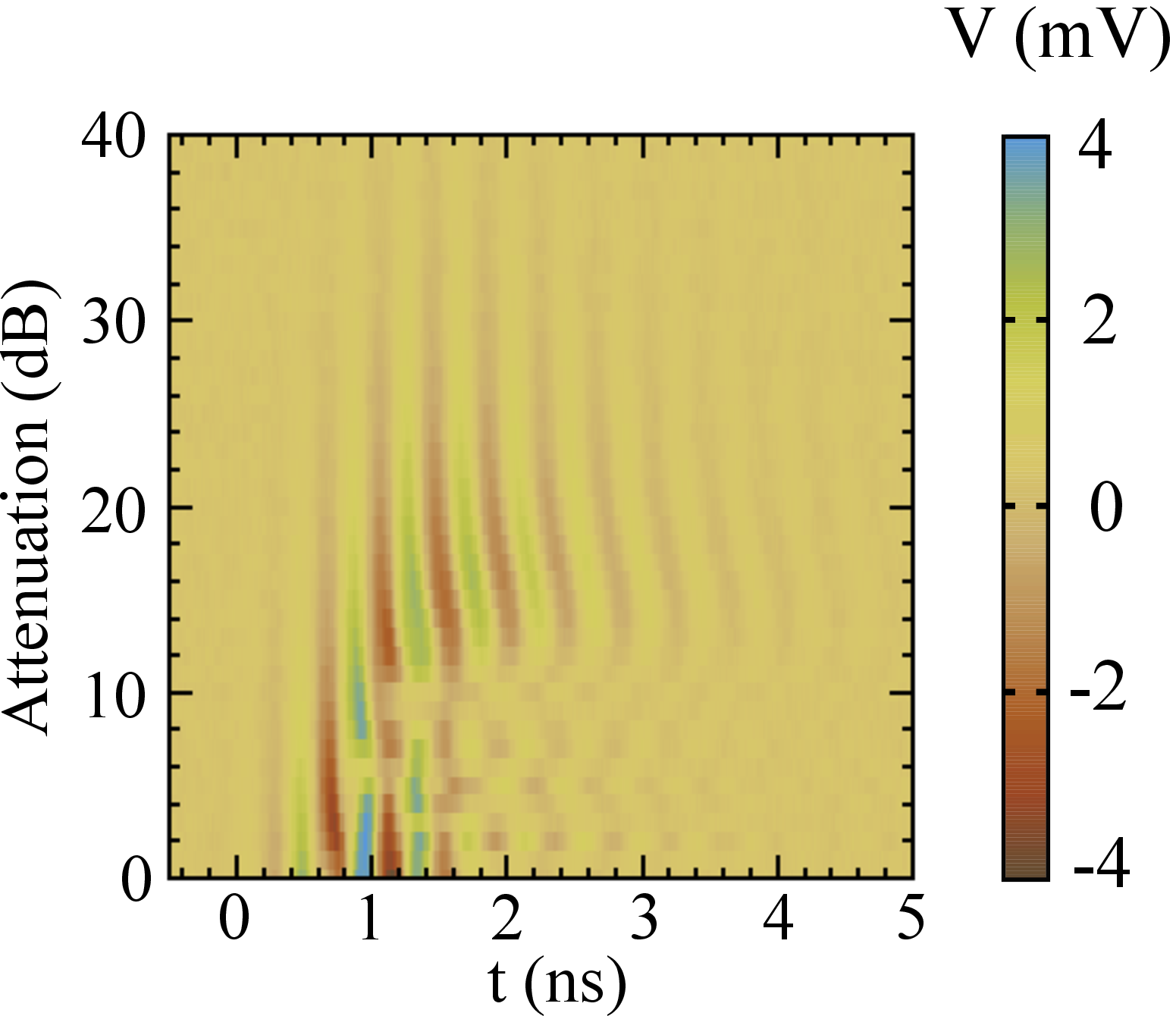}} }
  \caption{(Color online) A comparison of the amplitude dependence of ringdowns at a non-integer flux bias of $f_{s}=0.30$, for the input signal of Fig.~\ref{fig:single-ringdown}(a) at $2.4\,{\rm GHz}$. The plot in (a) shows experimental data, while in (b), the corresponding simulations. The attenuation on the drive pulse in dB is shown on the y-axis, with decreasing numbers implying increasing amplitude of the input pulse. The ringdown time is shown on the x-axis while the ringdown amplitude is represented by the color scale. The purple dashed line in (a) indicates the amplitude corresponding to the ringdown trace shown in Fig.~\ref{fig:single-ringdown}(b). }
    \label{fig:ampScansComparison1ns25GHz}
\end{figure}

\subsection{Amplitude Scans}
\label{ssec:AmplitudeScans}
We first look at amplitude scans, where we fix the applied flux bias $f_{s}$ and vary the amplitude of the microwave burst, recording a ringdown trace for each excitation strength of the SQUID oscillator. Figure~\ref{fig:ampScansComparison1ns25GHz} displays density plots of such a case, obtained with experimental data (top plot) and from simulations (bottom plot). The flux $f_{s}$ is fixed at $0.30$, which corresponds to the natural frequency of the SQUID of $2.4\,{\rm GHz}$ (when the SQUID is operated in a linear regime), satisfying a resonance condition. The y-axis, is the attenuation setting from the highest ($40\,{\rm dB}$) to the lowest attenuation ($0\,{\rm dB}$) corresponding to increasing burst amplitude towards the bottom of the plot and the x-axis is the ringdown time in nanoseconds. The color scale indicates the amplitude of the ringdown.
With increasing pulse strength, the frequency of the ringdowns decreases. This becomes particularly pronounced for attenuation levels less than $20\,{\rm dB}$. This shift to lower frequencies arises because the fictitious particle, whose position coordinate can be described by the $\phi_2$ degree of freedom, begins to explore the nonlinear (flatter) part of the potential energy landscape. For attenuation levels less than $12\,{\rm dB}$, we observe a sharp drop in the resulting ringdowns. At this point, the strength of the drive is now of the order of the critical current of the SQUID. The stochastic nature of the thermal noise causes different realizations to escape the potential well at different times, which in turn causes a substantial decay in the ringdown signal strength. This phenomenon is discussed in more detail in the next section.

\subsection{Escape from the Potential Well}
\label{ssec:escape}

\begin{figure*}[tp]
  \centering
  \captionsetup[subfigure]{position=top,singlelinecheck=off,topadjust=-5pt,justification=raggedright}
  \subfloat[]{       \includegraphics[width=0.32\textwidth]{{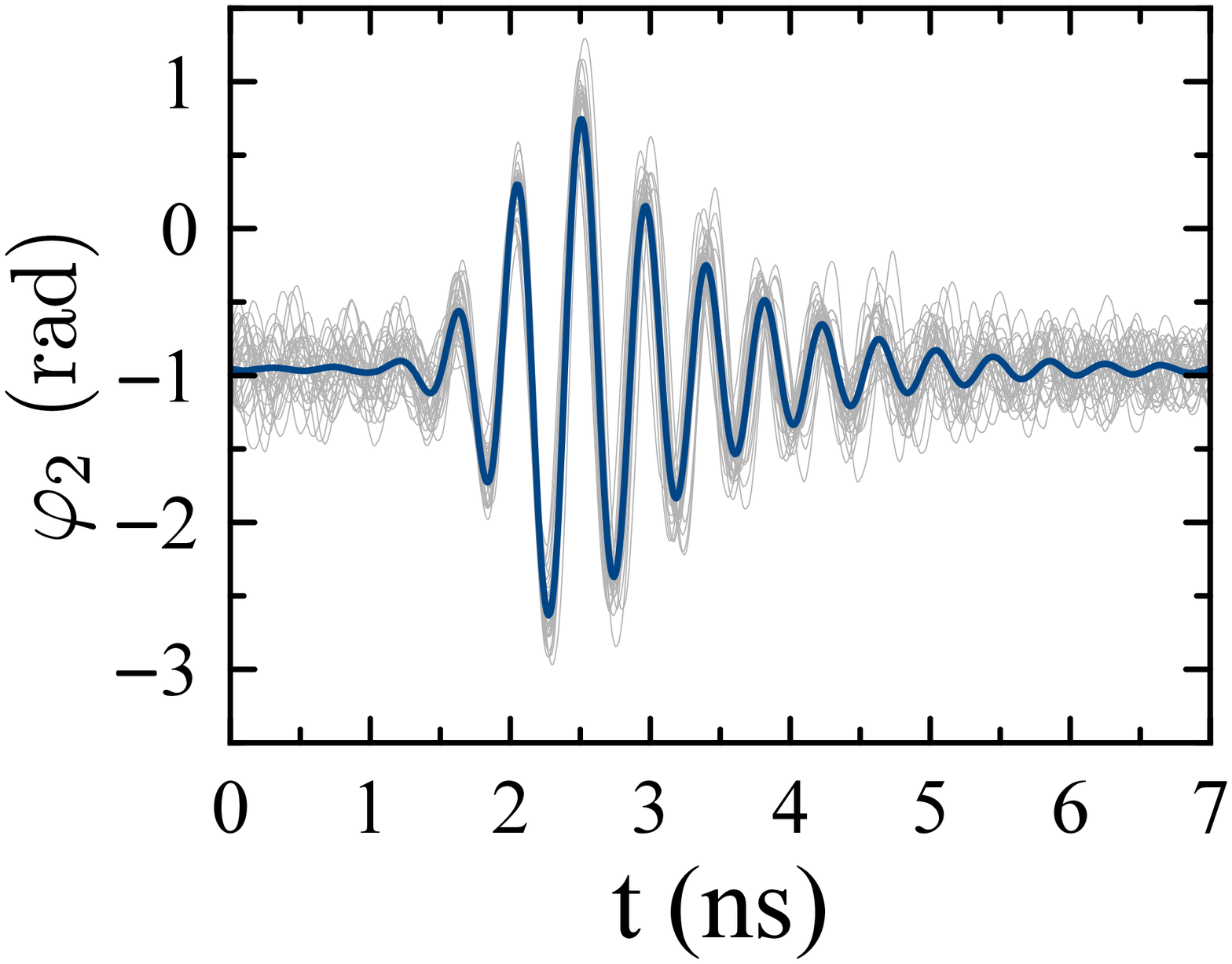}}  }
  \subfloat[]{       \includegraphics[width=0.32\textwidth]{{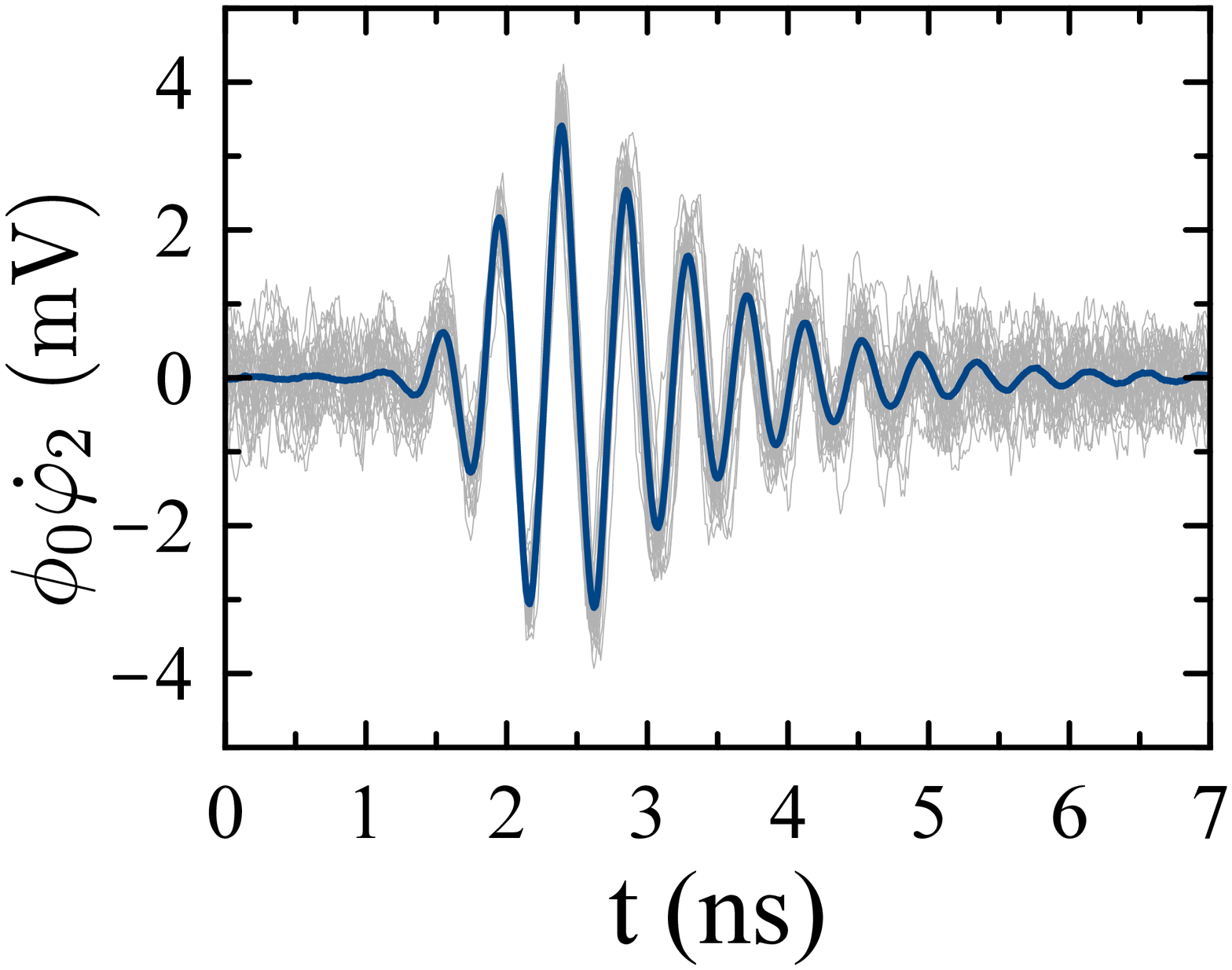}}  }
  \subfloat[]{       \includegraphics[width=0.32\textwidth]{{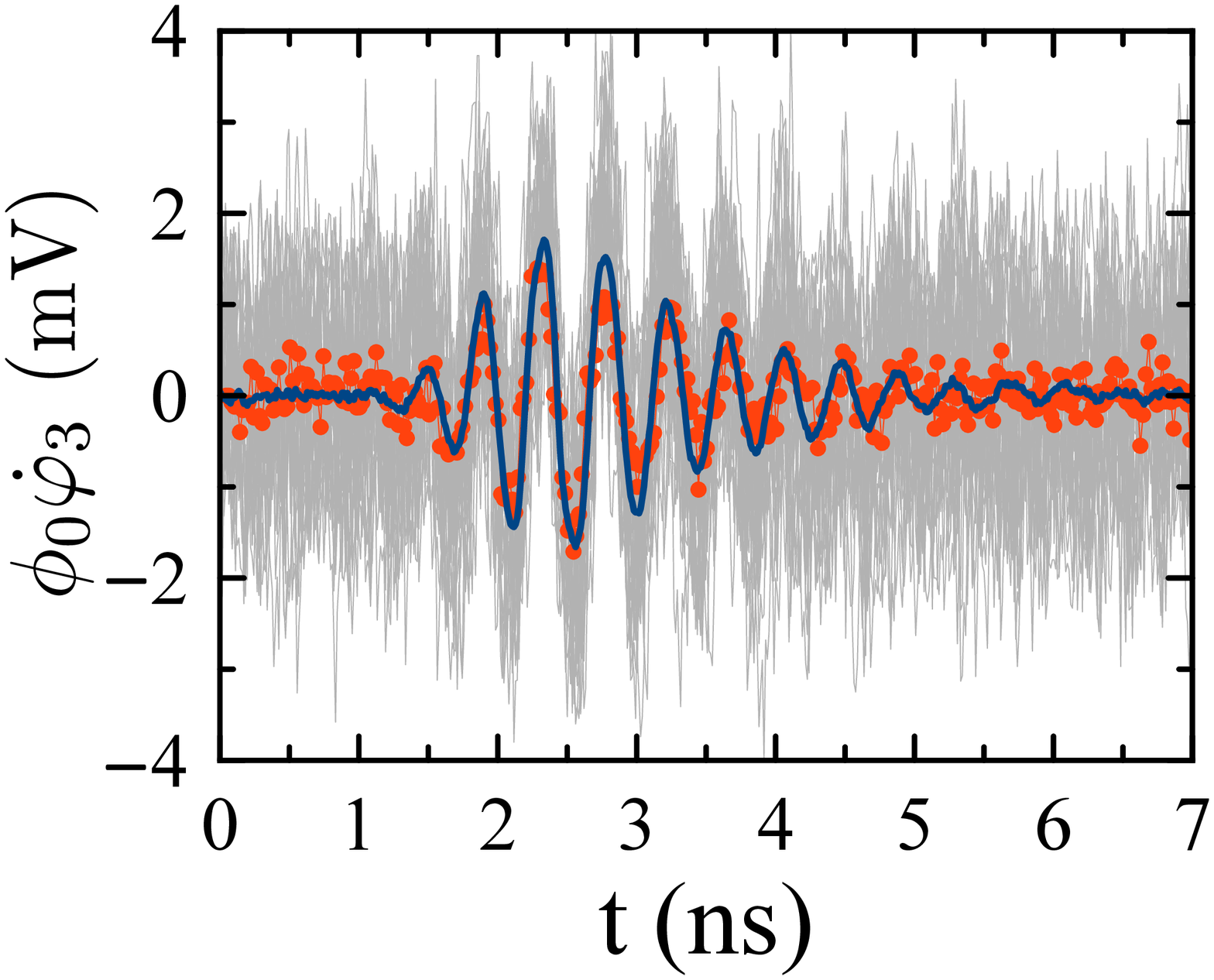}}  } \\
  \vspace{-0.4cm}
  \subfloat[]{       \includegraphics[width=0.32\textwidth]{{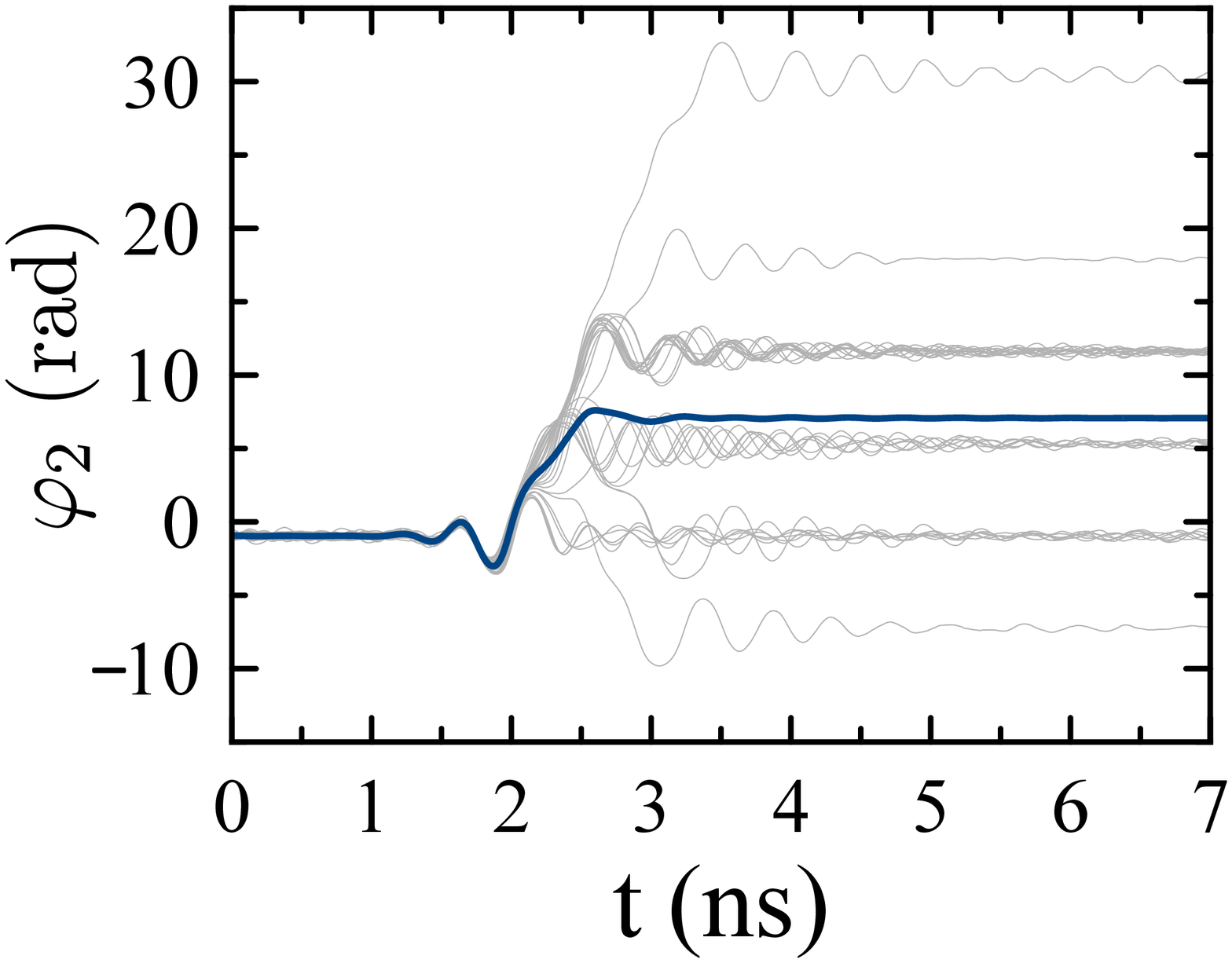}}  }
  \subfloat[]{       \includegraphics[width=0.32\textwidth]{{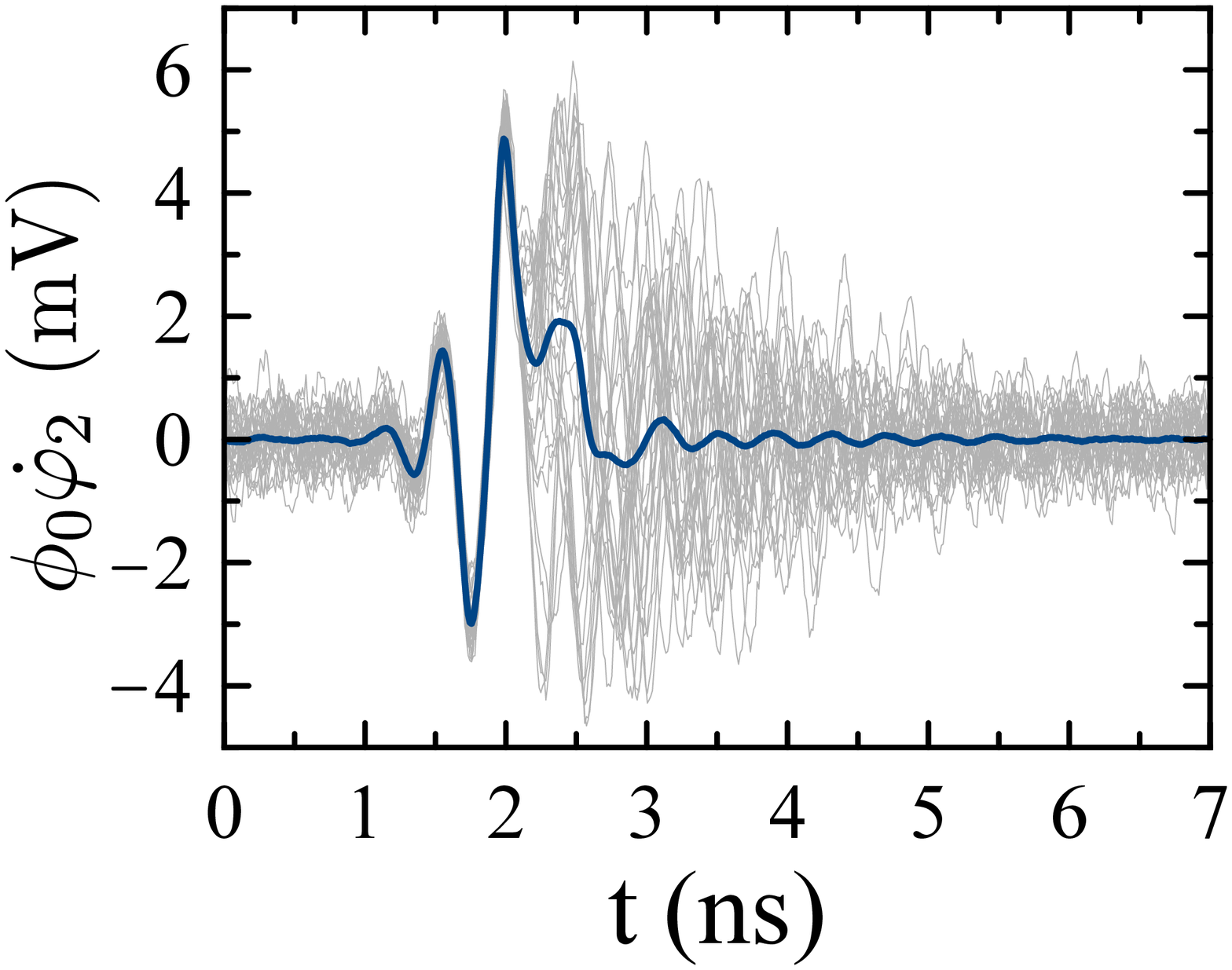}}  }
  \subfloat[]{       \includegraphics[width=0.32\textwidth]{{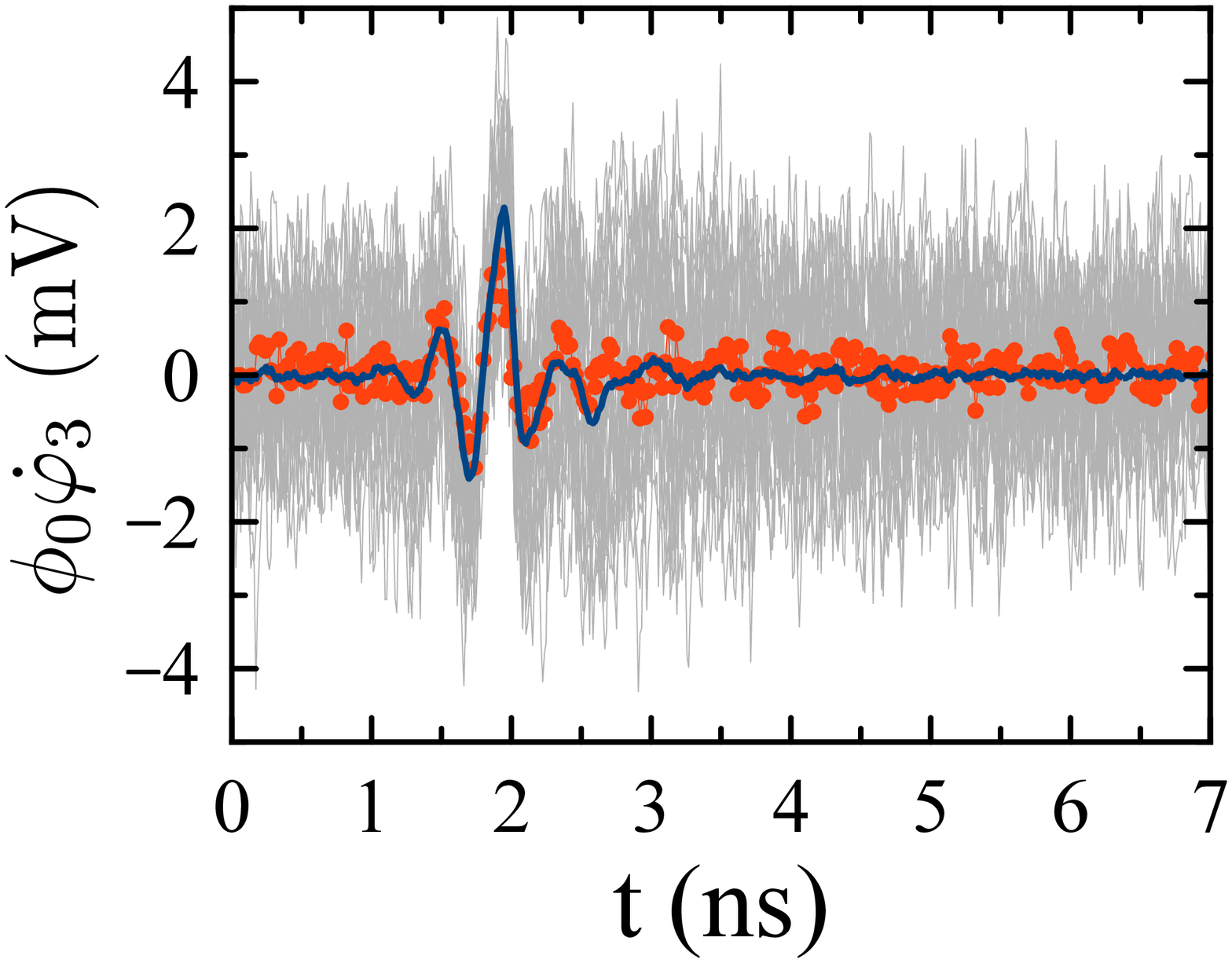}}  }
        \caption{(Color online) Low and high drive response. Each row shows plots of $\varphi_{2}(t)$, $ \phi_{0} \dot \varphi_{2}(t)$ and $ \phi_{0} \dot \varphi_{3}(t)$ (the output voltage) respectively. In plots (a-c), the amplitude of the input pulse is low with respect to the critical current of the Josephson junctions, whereas in (d-f) it is high. The gray curves show a small subset of individual realizations used to calculate the averages (blue curves). In the case of the third column we also show the experimental data for the same parameters (red dots) and how it compares to the simulation results. From the plots, one clearly sees how when the drive is low, all the realizations stay within the same potential well (see plot a), and their relative phase shift only varies slightly. In the case of high drive, we observe that different realizations tend to end up in different wells at different times (as shown in plot d), which introduces a relative phase shift between them. This in turn leads to faster decay of the average voltage, which is reflected in plots (e) and (f).}
    \label{fig:ampScanEscape}
\end{figure*}


The problem of particle escape from a potential well due to thermal noise has been investigated both theoretically as well as experimentally in a variety of studies \cite{kramers1940brownian,ben1983thermal,han1992effect,Hanggi90,gronbech2004microwave}. 
In the case of a dc SQUID, this rate can be approximated to be proportional to $ \Omega \exp \left( - U_{b}/k_{B} T \right)$  where $U_{b}$ represents the potential energy barrier height that the particle has to overcome, and $\Omega$ the natural frequency along the direction of escape. In our case, since we do not ``tilt'' the potential with a dc bias current, the escape time (inverse rate) can be shown to be much larger than the typical experimental run time. This is true over almost all settings of the applied flux bias $f_{s}$, except when $f_{s} \simeq 0.50$, where the potential barrier is close to being flat.

Thermal fluctuations, however, still end up playing an important role in the evolution of the system. In particular, we find that during strong pulses that excite the system to amplitudes in the vicinity of the SQUID's critical current, the thermal fluctuations can cause a strong mixing in the phases of various realizations, resulting in a damping of the ringdowns. To our knowledge, no detailed analytical study of this effect, with strongly time-dependent, transient pulses has been performed yet, but we can still study the situation numerically. To do this, we once again fix the applied flux bias at $f_{s}=0.30$ as in the previous section, and concentrate on two different pulses: the first at an attenuation of $15\,{\rm dB}$ and the other at the attenuation of $10\,{\rm dB}$. From Fig.~\ref{fig:ampScansComparison1ns25GHz}, we can see that these correspond to cases where substantial ringdown voltage is observed (the former case) and where the ringdowns are dramatically suppressed (the latter case).

In order to understand this behavior in more detail, we look at the evolution of the individual realizations, that so far have been averaged to obtain results comparable with the experiment. We stress, however, that while the behavior of the full circuit is largely governed by the dynamics of the SQUID, the experiment only provides us access to the external voltage --- the voltage at node $3$ in Fig.~\ref{fig:circuit-model} --- which in our simulations is represented mathematically as $\phi_0 \dot \varphi_{3}$. To directly observe the stochastic nature of the escape from the potential well, we need to look at the individual realizations of the full system. 
Of particular interest are the following simulation variables: $\varphi_{2}$, which represents the phase (i.e., the effective ``position'') of the SQUID degree of freedom, that dominates the evolution of the system, $\phi_{0} \dot \varphi_{2}$, which represents the voltage across the SQUID (or alternatively an effective ``velocity'' of the particle in the well), and finally $\phi_{0} \dot \varphi_{3}$, which is the voltage that we can directly compare to the experimental data. Figure~\ref{fig:ampScanEscape} shows plots that describe the evolution of these variables as a function of time. The top row shows data for a case of the low amplitude, $15\,{\rm dB}$ attenuation pulse, while the bottom row shows the case of high amplitude, $10\,{\rm dB}$ attenuation pulse. The leftmost column represents $\varphi_{2}(t)$, the middle column $\phi_0 \dot \varphi_{2}(t)$, and finally the rightmost column is the output voltage, namely $\phi_{0} \dot \varphi_{3}(t)$. In each case, the grey curves represent a subset of realizations that are averaged (curves in blue). The red dots in the plots from the rightmost column represent experimental data for the same set of parameters as the simulations. 
The key signature of the escape can be seen in the leftmost column. Here, when the pulse amplitude is low (top row), virtually all the realizations stay within the same potential well --- as one can see by noting that they all oscillate around the same value of $\varphi_{2}$, namely $\varphi_{2} \sim - 0.30 \pi$. In the case of the high amplitude pulse (bottom row), different realizations jump out to different potential wells. The stochastic nature of the noise causes these jumps to happen at different times, which  leads to a randomly shifted phase, as well as a different steady state value of $\varphi_{2}$. This has a substantial effect on the ``velocity'' (or $\dot \varphi_{2}$) of these realizations, as shown in the central column of Fig.~\ref{fig:ampScanEscape}. The result is a dramatic randomization in the phase of $\dot \varphi_{2}$, and as a result, of $\dot \varphi_{3}$, which is proportional to the output voltage of the circuit. As we see from the experimental voltage (red dots), the agreement of the measured data with the simulations is good. Finally, we stress that including the stochastic effects of thermal noise in our simulations has been crucial in reproducing this behavior.

\begin{figure*}[tp]
\centering
  \includegraphics[width=\textwidth]{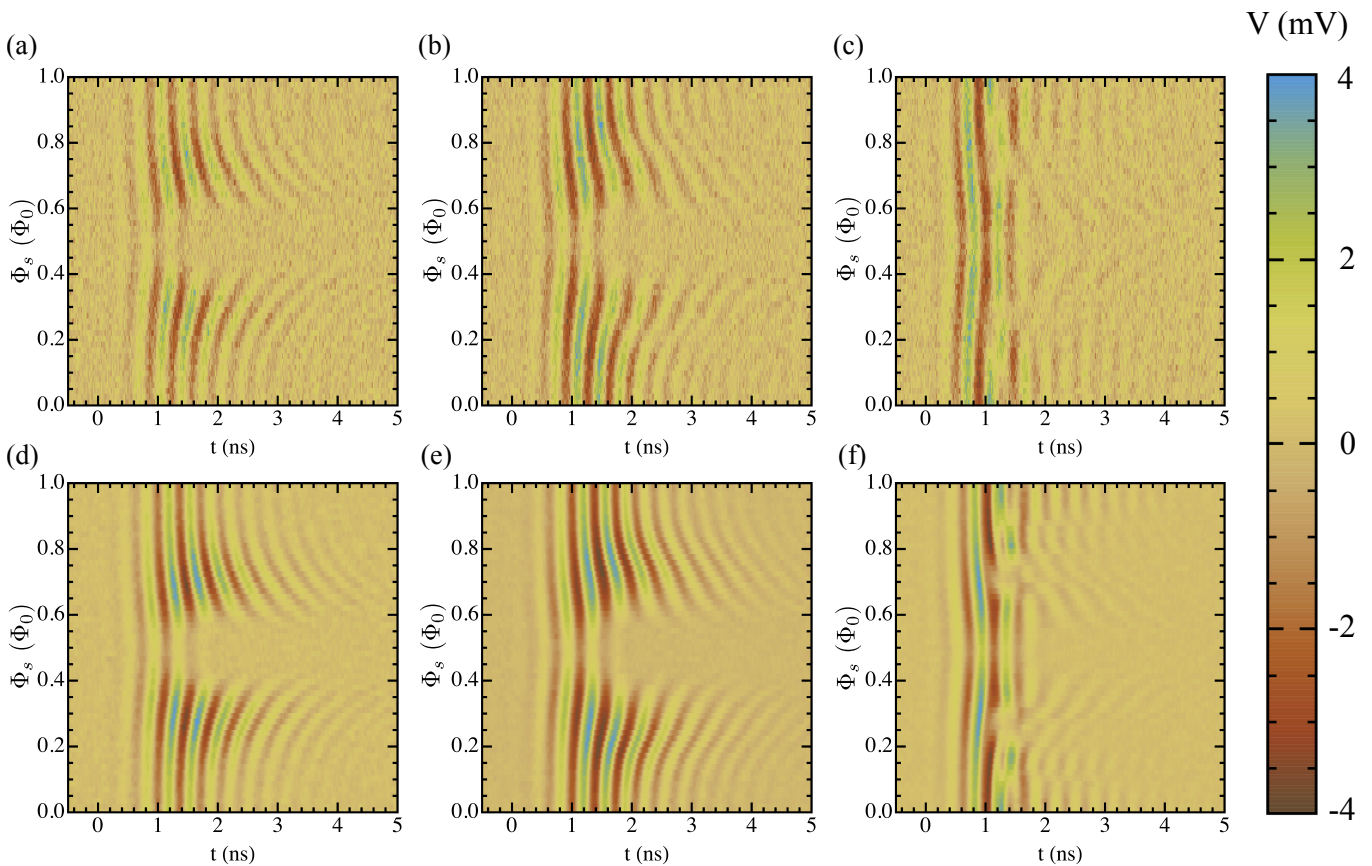}
  \caption{(Color online) Fixed amplitude flux scans with the applied flux bias $f_{s}$ between $0$ and $1$, and with the input signal of Fig.~\ref{fig:single-ringdown}(a) at $2.4$GHz. The top row (a-c) shows experimental data, while the bottom row (d-f), uses data obtained by running stochastic simulations. The amplitude of the input pulse increases from left to right, with the leftmost column showing results for $20\,{\rm dB}$ attenuation pulses, the middle column for $15\,{\rm dB}$ attenuation pulses, and finally the rightmost column for $10\,{\rm dB}$ attenuation pulses. As the amplitude increases, one clearly observes the effects of the nonlinearity of the system. See main text for a more detailed discussion.}
  \label{fig:fluxScans1ns25GHz}
\end{figure*}

\subsection{Flux Scans}
\label{ssec:fluxscan_theory}
We explore the voltage ringdown behavior further by studying their dependence on the magnetic flux applied to the SQUID. Here, the amplitude and frequency of the microwave burst is fixed while we vary the flux applied to the SQUID through one period of a flux quantum. The pulse frequency once again is chosen to correspond to resonance at the SQUID flux bias $f_{s} \sim 0.30$. The density plots of the flux-modulated ringdown traces is shown in Fig.~\ref{fig:fluxScans1ns25GHz} for three different pulse amplitudes. The top row shows plots obtained from experimental data, while the bottom row shows the simulations. The leftmost column has a low input pulse amplitude with $20\,{\rm dB}$ of attenuation, well below the critical current of the SQUID, the central column shows data for an input pulse with $15\,{\rm dB}$ of attenuation, while the rightmost column has a high amplitude pulse with $10\,{\rm dB}$ of attenuation.
By varying the applied flux through the SQUID, we are changing its effective inductance, and hence its natural frequency. 
It is worth stressing that this nonlinear dependence of the natural frequency on the applied flux is true even in the limit of small oscillations of the SQUID (where $|\varphi_{2}| \ll 1$), as was already discussed in Sec.~\ref{sec:MeasOfSquidOscillator}. Let us first concentrate on the leftmost column of Fig.~\ref{fig:fluxScans1ns25GHz}. Here the drive amplitude is still small and the nonlinearity of the potential energy in $\varphi_{2}$ is only beginning to play a role.
Yet, as the applied flux bias $f_{s}$ varies between 0 and 1, the ringdowns tend to fan out. As expected, the amplitude is largest near the flux bias of $f_{s} \sim 0.30$, since this is where the SQUID is resonant with the input pulse, and it is suppressed elsewhere. The results are also consistent with the fact that the natural frequency, up to zeroth order in $\beta$, is proportional to $\sqrt{\cos(\pi f_{s})}$. Hence, near $f_{s}\sim 0$, the variations in the ringdown structure are small, while at the same time, one sees a very abrupt suppression near $f_{s}\sim 0.50$. Here the effective natural frequency of the SQUID is very small and the short input pulse is unable to induce strong oscillations. This is an adiabatic regime, where the excitation of the circuit strongly follows the input pulse. In this regime, the ringdown suppression due to a highly off-resonant pulse can be confirmed further by studying individual realizations and showing that they stay in the same potential energy well as they started in, in contrast to what is observed during an escape --- see Sec.~\ref{ssec:escape}. Furthermore, a very similar ringdown structure can be obtained in a case where a simple harmonic oscillator, with the same flux-dependent form of natural frequency, is driven with the same pulse waveform. 
The situation is largely similar in the middle column of Fig.~\ref{fig:fluxScans1ns25GHz}. The key difference here is that now, not only is the natural frequency of the system nonlinear in the applied flux, but the amplitude of the input pulse is large enough for the SQUID degree of freedom $\varphi_{2}$ to start exploring the nonlinear regions of the potential energy well. This effect is particularly strong around the applied flux bias for which the SQUID is resonant with the input pulse (near $f_{s}\sim 0.30$). This in turn affects the degree of variation of the ringdown frequency with respect to $f_{s}$, as can be seen in the plots.
Finally, in the rightmost column we see a case of a strongly driven system. The resulting plots show an overall suppression of ringdown oscillations across all values of $f_{s}$, when compared to the instances with smaller drive amplitudes. In this case, the reason is two-fold. By once again studying the individual realizations as in Sec.~\ref{ssec:escape}, we can conclude that for the applied flux away from $f_{s}=0.50$, the main culprit in the suppression is the randomization of the phase of $\varphi_{2}$ due to the stochastic escape from the potential well. Near $f_{s}=0.50$ however, as in the case of low amplitude pulses, the main reason for the suppression is the off-resonance condition, where the frequency of the pulse is much greater that the natural frequency of the SQUID.
\section{Application to Flux Measurements}
\label{sec:Applications}
The high sensitivity of SQUIDs to applied flux has made them exquisite detectors of flux signals. They have been useful in various metrology experiments \cite{SQUID-Handbook,clarke2010squids,mcdermott2004microtesla,lanting2005frequency,drung2007highly}
 and more recently have played an important role in the field of quantum computing, as measurement devices for flux qubits \cite{chiorescu2004, lupascu2007,vion2002, claudon2007,van2003engineering}. In these applications, a flux qubit is typically coupled inductively to a SQUID and hence affects the net applied flux that is threaded through the device. 
 One of the original SQUID-based readout approaches involves biasing the SQUID with an appropriately selected dc current, such that the SQUID is put in a running state with non-zero voltage, with a high probability if the qubit is in one state, and with negligible probability when the qubit is in the other state \cite{vanderwal2000}. In an alternate approach, the SQUID is driven with a continuous sinusoidal signal while monitoring the resulting phase shift, which is qubit-state dependent \cite{lupacscu2006high}. Yet another proposed readout scheme uses brief, but strong, current bias pulses to the SQUID, which result in ringdown dynamics with an amplitude, and possibly phase, that depend on the qubit state \cite{serban2008}. This is somewhat analogous to what is presented in the experiment described here, although clearly in our case the flux differences are due to a global flux biasing as no qubit is actually present. Furthermore, the theoretical proposal outlined in \cite{serban2008} shows a full quantum treatment of the qubit and SQUID system, but only considers an ultra-short dc pulse, much shorter than the inverse characteristic qubit frequency. Due to the relatively high temperature of our measurements, our present experiment is in the classical regime, and involves input pulses with a time scale comparable to the dynamics of the circuit. 
\begin{figure}[tp]
\centering
  \includegraphics[width=0.36\textwidth]{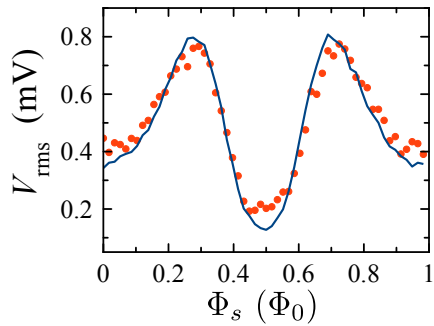} 
  \caption{(Color online) Root mean square output voltage $V_{\rm rms}$ as a function of the applied flux bias $\Phi_{s}$, calculated over a time range between 2.1 and $3.4\,{\rm ns}$. The plot uses data obtained with the $20\,{\rm dB}$ attenuation input pulses, and is the same as in the left column of the flux-scans from Fig.~\ref{fig:fluxScans1ns25GHz}. The dots represent results calculated from the experimental data (top row, leftmost column), while the solid line is produced using the simulations (bottom row, leftmost column). By biasing the flux through the SQUID near a point where the slope is high, for example at $\Phi_{s} \sim 0.36 \Phi_{0}$, one can have a means of distinguishing between different flux signals --- see main text for more details.}
\label{fig:rmsVoltage}
\end{figure}
Nevertheless, it is still useful to explore briefly just how the process of discrimination between two or more different flux states could be accomplished with our system. First, one can expect that the total applied flux through the SQUID would consist of some static bias flux $\Phi_{\rm bias}$ plus a signal flux that is to be measured, say $\Phi_{\rm signal}$.
One could then send a microwave pulse through the SQUID, analogous to what was considered here, and record the corresponding ringdown voltage. Some post-processing of this ringdown voltage, such as, for example, taking its root-mean-squared value, $V_{\rm rms}$, integrated over a suitably chosen time range, would provide a level corresponding to the signal flux. As long as these levels of various values of $\Phi_{\rm signal}$ can be distinguished, one has an effective flux meter. For a given input pulse, assuming that $V_{\rm rms}$ is a well behaved function of the total SQUID flux $\Phi_s$, one possible way to find the best $\Phi_{\rm bias}$ over a range of flux where $V_{\rm rms}$ is monotonic would be simply to look for the largest slope of $V_{\rm rms}$ with respect to $\Phi_{s}$, namely maximizing $\partial V_{\rm rms}(t_{\rm int}, \Phi_{s}) / \partial \Phi_{s}$ over all possible flux $\Phi_{s}$ between 0 and $0.5\Phi_{0}$ (due to symmetry) and integration times $t_{\rm int}$. This would give the greatest contrast between the cases of $\Phi_{\rm bias}+\Phi_{\rm signal}$ and $\Phi_{\rm bias} - \Phi_{\rm signal}$. 
Figure \ref{fig:rmsVoltage} shows an explicit example of this kind of flux discrimination based on our measurements, where we calculate $V_{\rm rms}$ over a time range between 2.1 and $3.4\,{\rm ns}$. The data that is being used corresponds to the input signal with $20\,{\rm dB}$ attenuation, and is the same as in the flux-scans from Fig.~\ref{fig:fluxScans1ns25GHz}. The dots represent results obtained from the experimental data (top row, leftmost column), while the solid line is produced from the evolution calculated through stochastic simulations (bottom row, leftmost column).
We can further make a crude calculation of the required sensitivity that one would need with the data from Fig.~\ref{fig:rmsVoltage} to distinguish between two hypothetical flux qubit states. Setting the bias flux at $\Phi_{\rm bias} \sim 0.36 \Phi_{0}$ the slope is roughly $5\, {\rm mV}/\Phi_{0}$. If we assume a conservative noise temperature of $150\,{\rm mK}$ for a $\sim 3\,{\rm GHz}$ amplifier with a bandwidth of $100\,{\rm MHz}$ \cite{defeo2012,ribeill2011}, the rms voltage noise at the amplifier input would be $\sim 200\, {\rm nV}$. If we take the $5\,{\rm mV}/ \Phi_{0}$ slope for the signal at the output extracted from Fig.~\ref{fig:rmsVoltage}, and divide by the net gain of the HEMT amplifiers $(\sim 55\,{\rm dB})$, this becomes $9\,{\rm \mu V}/\Phi_{0}$ at the SQUID oscillator output. We consider a peak-to-peak qubit flux signal of $22\,{\rm m}\Phi_{0}$, which is reasonable\cite{vanderwal2000,PRB051,plourde2005flux}, considering the back-action on the qubit would also likely be significantly less compared to a switching dc SQUID measurement, since the SQUID never enters the running state. This then corresponds to a SNR of $\sim 1$. So, we would be right at the threshold for reading out the ringdowns and distinguishing between the two qubit states in a single shot. 
We should further stress that one could likely do better by both using more sensitive amplifiers and optimizing various parameters. Of particular importance would be integration time $t_{\rm int}$, the pulse amplitude, as well as the quality factor of the SQUID oscillator, all of which the $V_{\rm rms}$ curves are highly dependent on. 

\section{Conclusions} 
\label{sec:Conclusions}
In conclusion, we have studied the transient behavior of a dc SQUID operated as a nonlinear oscillator under pulsed ac excitation. Both experimentally as well as numerically, we applied signals of various amplitudes for different flux bias, while observing the resulting voltage ringdowns. In order to account for the non-zero temperature of the experiment, we used the Johnson-Nyquist approach and modeled resistors as noisy current sources. This let us numerically reproduce the stochastic escape dynamics observed when the SQUID was driven with high-amplitude pulses. Finally, we briefly discussed the potential applicability of our system, and in particular the observed ringdown dynamics, to flux measurement. We found a good general agreement between the experimental data and results obtained through numerical simulations.

\section{Acknowledgments} 
\label{sec:Acknowledgments}
This work was supported by the DARPA/MTO QuEST program through a grant from AFOSR. PG and FKW acknowledge support by NSERC and FKW also support by the European union through SCALEQIT. 
Device fabrication was performed at the Cornell NanoScale Facility, a member of the National Nanotechnology Infrastructure Network, which is supported by the National Science Foundation (Grant ECCS-0335765). PG would like to thank Chunqing Deng for a useful discussion on the Born-Oppenheimer approximation, and Anthony J. Leggett for his helpful remarks regarding the applicability of our model.

\appendix
\section{Equations Of Motion}
\label{sec:EquationsOfMotion}
In this section we present a derivation of the equations of motion of a circuit which was used to model our experimental apparatus. We start with a full description of the system at zero-temperature and reduce the equations of motion by eliminating fast degrees of freedom. We then add the effects of the temperature-dependent noise.
\\
\subsection{Zero Temperature}
\label{ssec:ZeroTemperature}
As discussed in Section~\ref{sec:ExperimentalSetup} above, the circuit diagram is shown in Fig.~\ref{fig:circuit-model}. Our model assumes that the external flux is delivered directly to the SQUID loop, and other branches have no intrinsic geometric inductance. We further neglect the mutual inductance in the system other than the one that mediates the external flux $\Phi_{s}$. To obtain the equations of motion, we follow the treatment of Devoret \cite{Devoret95}. With each node $i$, we associate a corresponding node flux $\Phi_{i}$ related to a node voltage by $\Phi_{i} = \int_{-\infty}^{t} dt' V(t')$. We express the currents across elements in terms of $\Phi_{i}$ and using Kirchoff's current conservation conditions at each node $i$, arrive at the equations of motion
\begin{align}
    \begin{split}
    \frac{1}{R_z} ( V_{\rm{in}} - \dot \Phi_{1} ) =&   C_{\rm{in}} ( \ddot \Phi_{1} - \ddot \Phi_{2} )  \\
    C_{\rm{in}} ( \ddot \Phi_{1} - \ddot \Phi_{2} ) =&  \frac{2}{L_{g}} ( \Phi_{2} - \Phi_{4} + \Phi_{s}) + \frac{2}{L_{g}} ( \Phi_{2} - \Phi_{5}) \nonumber \\
    &+ C_{t} \ddot \Phi_{2} + \frac{1}{R_{t}} \dot \Phi_{2}  + C_{\rm{out}} (\ddot \Phi_{2} - \ddot  \Phi_{3} )  \\ 
    C_{\rm{out}} (\ddot \Phi_{2} - \ddot \Phi_{3} ) =&  \frac{1}{R_z} \dot \Phi_{3} \\
    \frac{2}{L_{g}}  (\Phi_{2} - \Phi_{4} + \Phi_{s}) =&  I_{0} \sin ( \Phi_{4}  2 \pi / \Phi_0 ) + \frac{1}{R_i} \dot \Phi_{4} + C_{J} \ddot \Phi_{4}  \\
    \frac{2}{L_{g}} (\Phi_{2} - \Phi_{5} ) =&  I_{0} \sin ( \Phi_{5} 2 \pi / \Phi_0 ) + \frac{1}{R_i} \dot \Phi_{5} + C_{J}\ddot \Phi_{5} 
    \end{split}
\end{align}
Next, taking the flux quantum $\Phi_0=2.07 \times 10^{-15} \text{Wb} = 2 \pi \phi_0$, we perform a change of variables so that $\Phi_{i}=\frac{\Phi_{0}}{2 \pi} \varphi_{i}=\phi_{0} \varphi_{i}$. Here, a difference $\varphi_{i} - \varphi_{j}$ for some $i\ne j$, corresponds to the superconducting phase difference. We further take $\varphi_{\pm}=\frac{1}{2}\left( \varphi_{4} \pm \varphi_{5} \right)$, $C_{\Sigma}=C_{\rm{in}} + C_{\rm{out}} + C_{t}$, $L_{J0}=\Phi_{0}/2 \pi I_0$ and rewrite the external flux $\Phi_{s}$ in terms of the ratio $f_{s}=\frac{\Phi_{s}}{ \Phi_{0}}$. After dividing all equations by $\phi_{0}$, we have
\begin{align}
    \begin{split}
0 =&     C_{\rm in} \ddot \varphi_{1} - C_{\rm in} \ddot \varphi_{2} +  \frac{1}{R_{z}} \dot \varphi_{1} - \frac{1}{\phi_0  R_{z} } V_{\rm{in}}  \\
0 =&    - C_{\rm{in}}  \ddot \varphi_{1} + C_{\Sigma} \ddot \varphi_{2} - C_{\rm{out}} \ddot \varphi_{3} + \frac{1}{ R_t} \dot \varphi_{2} \nonumber \\ 
 &+ \frac{4}{L_g  } \left( \varphi_{2} - \varphi_{+} + \pi f_{s} \right)  \\
0 =&    - C_{\rm out}  \ddot \varphi_{2} + C_{\rm out} \ddot \varphi_{3}  + \frac{1}{ R_z} \dot \varphi_{3}  \\
0 =&    2 C_{J} \ddot \varphi_{+} + \frac{2}{L_{J0}} \sin \varphi_{+} \cos \varphi_{-}  - \frac{4}{   L_{g}} \left(  \varphi_{2} -  \varphi_{+}     + \pi f_{s}   \right) \\
0 =&  2  C_{J}  \ddot \varphi_{-} + \frac{ 2 }{ L_{J0} }  \sin \varphi_{-} \cos \varphi_{+}  - \frac{4}{  L_{g}} \left( -  \varphi_{-}   + \pi f_{s} \right).
    \end{split}
\end{align}
Thus, we end up with equations of motion for five degrees of freedom. In order to further simplify the above, we note that in our case, the capacitances (or effective masses) of oscillators $\varphi_{+}$ and $\varphi_{-}$ are two orders of magnitude smaller than that of $\varphi_{2}$. Furthermore, the Josephson inductance $L_{J0}$ is much greater than the geometric inductance $L_{g}$. This allows us to apply an Oppenheimer-Born-like approximation and eliminate the fast-oscillating degrees of freedom $\varphi_{+}$ and $\varphi_{-}$. To do this, we first define a potential energy $U$ that can be associated with our system. Neglecting terms due to the external drive, we have
\begin{align}
    \frac{U}{2 E_{J}} =& -\cos \varphi_{+} \cos \varphi_{-}  + \frac{1}{\beta} \left( \varphi_{-} - \pi f_{s} \right)^{2} \nonumber \\
    &+ \frac{1}{\beta} \left( \varphi_{+} - (\varphi_{2} + \pi f_{s} ) \right)^{2},
    \label{eq:energyU}
\end{align}
with $\beta=L_{g}/L_{J0}$. Next, we fix the slow variable $\varphi_{2}$, and note that since $\beta \ll 1$, the second and third terms in Eq.~(\ref{eq:energyU}) will dominate. Hence the minima of $U$ will be close to $\varphi_{+}=\varphi_{2}+\pi f_{s}$ and $\varphi_{-}=\pi f_{s}$. By expanding $U$ near these points and minimizing, we can calculate the corrections to the minimum points. Keeping terms up to first order in $\beta$ we arrive at
\begin{align}
    \varphi_{-}^{\rm min} = \pi f_{s} - \beta \frac{\sin \left( \pi f_{s} \right) \cos \left( \pi f_{s} + \varphi_{2} \right)}{ 2 } \\
    \varphi_{+}^{\rm min} = \pi f_{s} + \varphi_{2} - \beta \frac{\cos \left( \pi f_{s} \right) \sin \left( \pi f_{s} + \varphi_{2} \right)}{ 2 }.
    \label{eq:dd}
\end{align}
These results are then substituted back into the expanded potential energy, which leads to $U_{\rm eff}=U_{0} + U_{1}$, now only in terms of $\varphi_{2}$ and with 
\begin{align}
    \begin{split}
        \frac{U_{0}}{2 E_{J}} =&  - \cos\left( \pi f_{s} \right)  \cos \left( \varphi_{2} + \pi f_{s} \right)   \nonumber 
    \label{eq:energyUeff}
    \end{split}
\end{align}
and 
\begin{align}
       \frac{U_{1}}{2 E_{J}} =& - \frac{\beta}{2} \left( \sin^{2}\left( \pi f_{s} \right) \cos^{2}(\pi f_{s} + \varphi_{2}) \right. \nonumber  \\
       & \left. +  \cos^{2}\left( \pi f_{s} \right) \sin^{2}(\pi f_{s} + \varphi_{2}) \right).
\end{align}
We have distinguished between the contributions to the effective potential energy between terms of different orders in $\beta$. $U_{0}$ neglects the geometric inductance completely, while $U_{1}$ shows the correction up to first order in $\beta$. 
We can hence write a set of effective equations of motion with $U_{\rm eff}$ as the potential energy, while now including drive term, as
\begin{align}
0 =&    C_{\rm in}  \ddot \varphi_{1}  - C_{\rm in} \ddot \varphi_{2} + \frac{1}{R_{z}} \dot \varphi_{1}- \frac{1}{ \phi_0  R_{z} } V_{\rm{in}}  \\
0 =&    - C_{\rm{in}}  \ddot \varphi_{1} + C_{\Sigma}  \ddot \varphi_{2} - C_{\rm{out}}  \ddot \varphi_{3} + \frac{1}{R_t} \dot \varphi_{2} \nonumber \\
&+ \frac{2}{  L_{J0}} \cos \left( \pi f_{s} \right) \sin \left( \varphi_{2} + \pi f_{s} \right) \nonumber   \\
&- \beta \left(\frac{\sin(4 \pi f_{s}+2 \varphi_{2})+\sin(2 \varphi_{2})}{2 L_{J0}} \right)  \\
   0 =&   - C_{\rm out} \ddot \varphi_{2} + C_{\rm out}  \ddot \varphi_{3}   + \frac{1}{ R_z} \dot \varphi_{3}.
    \label{eq:reducedSystem}
\end{align}
The next step is to account for the non-zero temperature of the system.
\subsection{Non-zero Temperature}
\label{ssec:Temperature}
We find that the accounting for thermal noise is of particular importance when comparing with the behavior of the experimental system in our simulations, in particular at high amplitude pulses. In order to model these effects, we use the thermodynamic dissipation-fluctuation relation \cite{nyquist1928thermal}.
 Thermal noise in the circuit is modeled by including a current noise source of strength $\sqrt{ \frac{2 k_{B} T }{ R_i}} n_{i}$ in parallel with each resistor $R_{i}$. We take $k_{B}$ as the Boltzman constant, and $T$ the temperature of the system. Furthermore, each $n_{i}(t)$ represents a normally distributed random variable, namely $n_{i} \in \mathcal{N}(0,1)$, that satisfies the following
\begin{align}
    \langle n_{i}(t) \rangle =&  0 \\
    \langle n_{i}(t) n_{j}(t') \rangle =&  \delta(t-t') \delta_{i,j}. 
    \label{eq:noise2}
\end{align}
Adding such a noisy current in parallel with each of the resistors to the effective model derived in Sec.~\ref{ssec:ZeroTemperature} leads to classical Langevin equations, which after rearranging, can be written in a vector form as Eq.~(\ref{eq:reducedSystem2Vec1}).

\bibliography{Oscillator}

\end{document}